\documentclass[12pt,preprint]{emulateapj}
\usepackage{epsfig}
\usepackage{epstopdf}
\usepackage{subfigure}

\begin{document}

\def\beq{\begin{equation}}
\def\eeq{\end{equation}}
\newcommand{\kms}{\,{\rm km\,s^{-1}}}
\newcommand{\msun}{\, M_\odot}
\newcommand{\lsun}{\, L_\odot}
\newcommand{\mlrsun}{\, M_{\odot} L_{\odot,R}^{-1}}
\newcommand{\mlvsun}{\, M_{\odot} L_{\odot,V}^{-1}}
\newcommand{\mbh}{M_\bullet}
\newcommand{\ml}{M_\star/L}
\newcommand{\mlr}{M_\star/L_R}
\newcommand{\mlv}{M_\star/L_V}
\newcommand{\rinf}{\, r_{\rm inf}}
\newcommand{\reff}{\,r_{\rm eff}}

\title{Dynamical Measurements of Black Hole Masses in Four Brightest Cluster Galaxies at 100 Mpc}

\author{Nicholas J. McConnell \footnotemark[1], Chung-Pei Ma \footnotemark[1], Jeremy D. Murphy \footnotemark[2], Karl Gebhardt \footnotemark[2], Tod R. Lauer \footnotemark[3], James R. Graham \footnotemark[1,4], Shelley A. Wright \footnotemark[4], and Douglas O. Richstone \footnotemark[5]}

\footnotetext[1]{Department of Astronomy, University of California at Berkeley, Berkeley, CA; nmcc@berkeley.edu, cpma@berkeley.edu, jrg@berkeley.edu}
\footnotetext[2]{Department of Astronomy, University of Texas at Austin, Austin, TX; murphy@astro.as.utexas.edu, gebhardt@astro.as.utexas.edu}
\footnotetext[3]{National Optical Astronomy Observatory, Tucson, AZ; lauer@noao.edu}
\footnotetext[4]{Dunlap Institute for Astronomy \& Astrophysics, University of Toronto, Toronto, Ontario; saw@di.utoronto.ca}
\footnotetext[5]{Department of Astronomy, University of Michigan at Ann Arbor, Ann Arbor, MI; dor@astro.lsa.umich.edu}

\begin{abstract}
We present stellar kinematics and orbit superposition models for the central regions of four Brightest Cluster Galaxies (BCGs), based upon integral-field spectroscopy at Gemini, Keck, and McDonald Observatories.  Our integral-field data span radii from $< 100$ pc to tens of kiloparsecs, comparable to the effective radius of each galaxy.  
We report black hole masses, $\mbh$, of $2.1^{+1.6}_{-1.6} \times 10^{10} \msun$ for NGC 4889, $9.7^{+3.0}_{-2.5} \times 10^9 \msun$ for NGC 3842, and $1.3^{+0.5}_{-0.4} \times 10^9 \msun$ for NGC 7768, with errors representing $68\%$ confidence limits.  For NGC 2832 we report an upper limit of $\mbh < 9.0 \times 10^9 \msun$.  
Our models of each galaxy include a dark matter halo, and we have tested the dependence of $\mbh$ on the model dark matter profile.  
Stellar orbits near the center of each galaxy are tangentially biased, on comparable spatial scales to the galaxies' photometric cores.  We find possible photometric and kinematic evidence for an eccentric torus of stars in NGC 4889, with a radius of nearly 1 kpc.   
We compare our measurements of $\mbh$ to the predicted black hole masses from various fits to the relations between $\mbh$ and stellar velocity dispersion ($\sigma$), luminosity ($L$), or stellar mass ($M_\star$).   
Still, the black holes in NGC 4889 and NGC 3842 are significantly more massive than all $\sigma$-based predictions and most $L$-based predictions.  The black hole in NGC 7768 is consistent with a broader range of predictions.  
\\

\end{abstract}

\pagestyle{plain}
\pagenumbering{arabic}

\maketitle

%
\section{Introduction}
\label{sec:intro}

For four decades, dynamical studies have provided evidence for the existence of black holes.  In the most exquisite cases, the orbital motions of individual stars or megamasers place strong lower limits on the density of a central object and rule out virtually all alternatives to a black hole
 \citep[e.g.,][]{Bolton72,Schodel02,Ghez05,Bender05,Herrnstein05}.
More generally, massive dark objects have been dynamically detected in a rapidly growing number of galactic nuclei and are widely assumed to be black holes 
(e.g., Sargent et al. 1978; Tonry 1987; for reviews and compilations see Kormendy \& Richstone 1995; 
G\"{u}ltekin et al 2009a; McConnell et al. 2011b).  
The menagerie of black holes spans from stellar-mass objects to ``supermassive'' behemoths, whose masses, $\mbh$, can approach $10^{10} \msun$.  Strong dynamical evidence for ``intermediate-mass'' black holes with $\mbh \sim 10^3 - 10^5 \msun$ is more elusive but has been recorded in a few stellar systems (Gebhardt et al. 2005; L\"{u}tzgendorf et al. 2011; Jalali et al. 2012; c.f. Baumgardt et al. 2003; van der Marel \& Anderson 2010).

In galactic nuclei, $\mbh$ appears to correlate with the stellar mass, luminosity, and velocity dispersion of the host bulge or spheroid \citep[e.g.,][]{Dressler89,KR95,Magorrian,Ferr00,Geb00,Tremaine02,MH03}.
A frontier goal is to explore the extrema of these black hole scaling relations by directly measuring $\mbh$ in low-mass galaxies and in extremely massive galaxies.  
Here we focus on Brightest Cluster Galaxies (BCGs), which are among the most massive galaxies in the present-day Universe.  BCGs typically reside deep in the gravitational potentials of rich clusters, and their environment may provide a unique path for galaxy and black hole growth.  BCGs have been observed to follow a steeper relation between luminosity ($L$) and velocity dispersion ($\sigma$) than less massive elliptical galaxies \citep[e.g.,][]{Bernardi07,Desroches,Lauer07,vdLinden}.  \citet{Bernardi07} and \citet{Lauer07} have noted that the $\mbh-L$ relation predicts systematically more massive black holes than $\mbh-\sigma$ for the most massive galaxies.

%
\begin{table*}[!t]
\begin{center}
\caption{Global properties of target galaxies}
\label{tab:sample}
\begin{tabular}[b]{cccccccccc}  
\hline
Galaxy & $D$ & $M_V$ & $\rinf$ & $\reff$ & $\sigma_{\rm eff}$ & $\;\;$ PA$_{\rm maj}$ & $b/a$ & $i$  & Features \\
& (Mpc) & & ($''$) & ($''$) & ($\kms$) & ($^\circ$) & & ($^\circ$) & \\
\\
(1) & (2) & (3) & (4) & (5) & (6) & (7) & (8) & (9) & (10) \\
\hline 
\\
NGC 4889 & 103.2 & -23.91 & 1.5 & 54.5 & 347 & 80 & 0.73 & 90 & \\
NGC 3842 & 98.4 & -23.21 & 1.2 & 37.8 & 270 & 5 & 0.86 & 90 & Isophotal twist at $r > 12''$\\
NGC 7768 & 112.8 & -22.92 & 0.14 & 23.1 & 257 & 60 & 0.75 & 90 & Nuclear dust disk\\
NGC 2832 & 101.9 & -23.87 & $< 0.7$ & 67.4 & 334 & -21 & 0.68 & 90 & Isophotal twist at $r < 8''$\\
\hline
\end{tabular}
\end{center}
\begin{small}
\textbf{Notes:}  Column 1: galaxy.  Column 2: co-moving distance, derived from cluster systemic velocity with $H_0 = 70 \kms$ Mpc$^{-1}$.
Column 3: absolute $V$-band magnitude.  Column 4: black hole radius of influence, derived from our best-fit value of $\mbh$ and the effective velocity dispersion in column 6.  Column 5: effective radius from \citet{Lauer07}.  Column 6: effective velocity dispersion, from luminosity-weighted measurements between $r_{\rm inf}$ and $r_{\rm eff}$.
Column 7: major-axis position angle, measured from north toward east.  This angle is used to define the equatorial plane of each galaxy in our stellar orbit models.   Column 8: observed axis ratio.  Column 9: inclination assumed for stellar orbit models.  $90^\circ$ corresponds to edge-on inclination.   Column 10: notable photometric features.
\end{small}
\vspace{0.15in}
\end{table*}

To date, there are eight groups or clusters where $\mbh$ has been measured dynamically in the massive central galaxy: Coma (NGC 4889; McConnell et al. 2011b), Fornax (NGC 1399; e.g. Houghton et al. 2006; Gebhardt et al. 2007), Virgo (M87; Gebhardt et al. 2011), Abell 1367 (NGC 3842; McConnell et al. 2011b), Abell 1836 (PGC 49940; Dalla Bont\`{a} et al. 2009), Abell 2162 (NGC 6086; McConnell et al. 2011a), Abell 3565 (IC 4296; Dalla Bont\`{a} et al. 2009) and the IC 1459 group (IC 1459; Cappellari et al. 2002).  Only three of these systems (Coma, Abell 1367, Abell 3565) are rich clusters.   
In Fornax and Virgo, the central cD galaxy is not even the brightest member.  The brightest galaxy in Fornax is NGC 1316, which lies near the cluster outskirts and hosts a black hole with $\mbh = 1.5 \times 10^8 \msun$ \citep{Nowak08}.  In Virgo, M49 anchors a sub-group more than 1 Mpc from the more centralized M87 and M86 sub-groups, and hosts a black hole with $\mbh = 1.5 \times 10^9 \msun$ (Shen et al., in prep.).  In order to thoroughly explore black hole and galaxy co-evolution in different environments, we must measure $\mbh$ in a larger sample of BCGs, with more examples from rich galaxy clusters.  
This paper includes expanded discussion of the black hole measurements in the BCGs of Coma and Abell 1367. 

This paper marks our continued effort to measure
$\mbh$ using stellar dynamics in BCGs beyond the Virgo cluster.
BCGs are rare objects and typically lie at large distances, making high-resolution observations difficult.
Additionally, their centers are typically fainter than less massive ellipticals, and 8- to 10-meter telescopes are required to obtain high-quality spectra at angular scales comparable to the black hole radius of influence,
$\rinf \equiv G\mbh\sigma^{-2}$.
For targets at distances $\sim 100$ Mpc and predicted black hole masses $\sim 10^9 \msun$ from the $\mbh-\sigma$ relation, adaptive optics (AO) is necessary to resolve $\rinf \sim 0.1''$.
Under good conditions, seeing-limited observations can resolve the gravitational influence of extremely massive black holes ($\mbh \sim 10^{10} \msun$).
Wide-field kinematic measurements are necessary to trace the galaxies' stellar mass profiles and dark matter halos.  

We report measurements of $\mbh$ and the $R$-band stellar mass-to-light ratio, $\mlr$, in four BCGs at distances of $\sim 100$ Mpc: NGC 4889 of the Coma cluster (Abell 1656), NGC 3842 (Abell 1367), NGC 7768 (Abell 2666), and NGC 2832 (Abell 779). 
We have obtained high-resolution data of the line-of-sight stellar velocities in the central regions of the four galaxies using instruments on the Gemini North and Keck Telescopes.  In addition, we have used the 2.7-meter telescope at McDonald Observatory to measure stellar kinematics at large radii.  At all spatial scales, we use integral-field spectrographs (IFSs) to obtain full 2-dimensional spatial coverage, which places tighter constraints on stellar orbits.  We determine $\mbh$ and $\mlr$ with axisymmetric orbit superposition models, which include a dark matter component in the gravitational potential.  
We have reported the black hole measurements of NGC 4889 and NGC 3842 in \citet{mcconnell11b}, but due to space limitations, only the basic information was presented there.  Here we provide a comprehensive discussion of the data analysis procedures, kinematic information, and stellar orbit modeling for all four galaxies.  

The paper is organized as follows.  
In \S2, we describe our photometric and spectroscopic observations and data reduction procedures.
In \S3, we describe our procedures for extracting two-dimensional kinematics from the data obtained on three IFSs (GMOS, OSIRIS, and the Mitchell Spectrograph).  The outputs of this step are non-parametric line-of-sight velocity distributions (LOSVDs) in two-dimensional spatial bins in the central regions of each galaxy.
In \S4, we show maps of the lowest four Gauss-Hermite moments of the LOSVDs for the four galaxies.  We discuss both the two-dimensional maps and radial profiles of the kinematic moments.
In \S5 we summarize the stellar orbit modeling procedure.  We report our measurements of $\mbh$ and $\mlr$ and describe how these measurements depend on the assumed dark matter halo profile, as well as other possible systematic biases.  We also discuss the relative contributions of radial and tangential orbits in each galaxy.
In \S6 we 
summarize our results and
compare our measurements of $\mbh$ to predictions from the $\mbh - \sigma$ and $\mbh - L$ relations. 
Throughout this paper, we assume $H_0 = 70$ km s$^{-1}$, $\Omega_m = 0.27$, and $\Omega_\Lambda = 0.73$.

%
\section{Observations}
\label{sec:obs}

\subsection{Global Galaxy Properties}
\label{sec:global}

%
\begin{figure*}[!t]
\centering
  \epsfig{figure=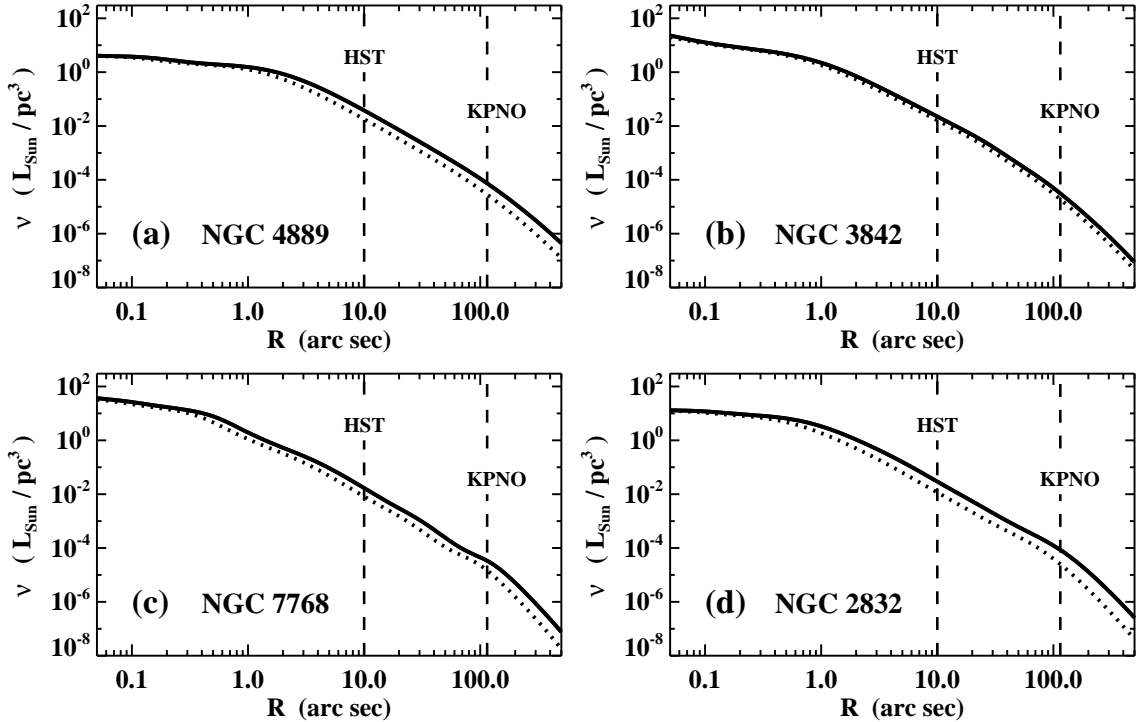,width=6.0in}
 \caption{De-projected $R$-band stellar luminosity density versus radius along the major axis (solid line) and minor axis (dotted line) of each galaxy.  The dashed vertical lines mark the outermost extents of photometric data from \textit{HST} and KPNO.  Luminosity densities beyond $115''$ are derived from a de Vaucouleurs surface brightness profile.  (a) NGC 4889.  (b) NGC 3842.  (c) NGC 7768.  (d) NGC 2832.}
\label{fig:lden}
\vspace{0.15in}
\end{figure*}

We list the basic properties of each galaxy in Table~\ref{tab:sample}.  We compute co-moving distances from the average velocity of the host galaxy cluster with respect to the cosmic microwave background (CMB).  We use heliocentric cluster velocities from \citet{LP94} and translate them to the CMB rest frame with the NASA/IPAC Extragalactic Database (NED).  
To compute the $V$-band luminosity, $L_V$, for NGC 4889, NGC 3842, and NGC 2832, we adjust the absolute magnitudes from \citet{Lauer07,Lauer07b} to our assumed distances.  For NGC 7768, we adopt the apparent $V$-band magnitude of 12.40 from the RC3 catalog \citep{deV91} and assume a luminosity distance of 115.8 Mpc.  

Galaxies with spatially resolved stellar kinematics are often placed on the $\mbh-\sigma$ relation using $\sigma_{\rm eff}$, the luminosity-weighted average velocity dispersion out to one effective radius ($\reff$).  Yet when the black hole radius of influence, $\rinf$, subtends a large angle, the conventional measurement of $\sigma_{\rm eff}$ can include a direct signal from the black hole.  In extreme cases, the velocity dispersion only reaches its ``average'' value at $r < \rinf$ (see, e.g., Gebhardt et al. 2011 for M87).  As the $\mbh-\sigma$ relation is widely interpreted as an empirical correlation between independent parameters, it is important to remove the black hole's direct influence on $\sigma_{\rm eff}$.  For our BCGs, we measure the luminosity-weighted average velocity dispersion for $\rinf \leq r \leq \reff$:  
\begin{equation}
\sigma_{\rm eff}^2 \equiv \frac{\int^{\reff}_{\rinf} \left(\sigma^2 + v_{\rm rad}^2 \right)I(r)dr}{\int^{\reff}_{\rinf} I(r)dr}  \;\; ,
\label{eq:sigeff}
\end{equation}
where $I(r)$ is the galaxy's one-dimensional stellar surface brightness profile.

We compute $\rinf$ and $\sigma_{\rm eff}$ from the integral-field kinematics and black hole masses presented herein; at each radius, we average $v_{\rm rad}$ and $\sigma$ over all polar angles sampled.  Because $\rinf \equiv G\mbh\sigma_{\rm eff}^{-2}$, we compute $\rinf$ and $\sigma_{\rm eff}$ iteratively; the iterations converge quickly in all cases.  For NGC 2832, we only have upper limits for $\mbh$ and $\rinf$, and we assume $\rinf = 0$.  For NGC 4889 and NGC 3842, our upper integration limit is the maximum radius of our kinematic data, which corresponds to $0.4 \reff$ and $0.8 \reff$, respectively.

%
\begin{table*}[!t]
\begin{center}
\caption{Spectroscopic observations}
\label{tab:spec}
\begin{tabular}[b]{llllllll}  
\hline
Galaxy & Instrument & Date & $t_{int}$ & $\lambda$ Range & PA & $\Delta x$ & FWHM\\
&  &  & (s) & (nm) & ($^\circ$) & ($''$) & ($''$)\\
\\
(1) & (2) & (3) & (4) & (5) & (6) & (7) & (8) \\
\hline 
\\
NGC 4889 & OSIRIS & May 2008 & $9 \times 900$ & 1473 - 1803 & 135 & 0.05 & \\
NGC 4889 & OSIRIS & May 2009 &  $9 \times 900$ & 1473 - 1803 & 80 & 0.05 &  \\
NGC 4889 & GMOS & March 2003 & $5 \times 1200$ & 744 - 948 & 90 & 0.2 & 0.4 - 0.7 \\
\\
NGC 3842 & OSIRIS & May 2010 & $9 \times 900$ & 1473 - 1803 & 10 & 0.05 & 0.06, 0.19 \\
NGC 3842 & GMOS & April 2003 & $5 \times 1200$ & 744 - 948 & -30 & 0.2 & 0.4 - 0.7 \\
NGC 3842 & Mitchell & March 2011 & $3 \times 1200$ & 358 - 588 & 0 & 4.1 &  \\
\\
NGC 7768 & OSIRIS & September 2010 & $14 \times 900$ & 1473-1803 & -125 & 0.05 & 0.04, 0.09  \\
NGC 7768 & Mitchell & September 2011 & $13 \times 1200$ & 358 - 588 & 0 & 4.1 &  \\
\\
NGC 2832 & GMOS & March 2003 & $4 \times 1200$ & 744 - 948 & 30 & 0.2 & 0.4 - 0.7 \\
NGC 2832 & Mitchell & January 2008 & $11 \times 1200$ & 354 - 584 & 0 & 4.1 &  \\
NGC 2832 & Mitchell & February 2008 & $28 \times 1200$ & 354 - 584 & 0 & 4.1 &  \\
\hline
\end{tabular}
\end{center}
\begin{small}
\textbf{Notes:}  Column 1: galaxy.  Column 2: instrument.  OSIRIS (OH-Suppressing Infra-Red Imaging Spectrograph) was used on Keck II with LGS-AO.  GMOS (Gemini Multi-Object Spectrograph) was used on Gemini North.  The Mitchell Spectrograph was used on the McDonald Observatory 2.7-m telescope.  Column 3: observing dates.  Column 4: number of science exposures $\times$ integration time per exposure.  
Column 5: observed wavelength range.  Column 6: position angle of the long axis for OSIRIS and GMOS, measured from north toward east.  Column 7: angular size of lenslets or fibers.  Column 8: PSF FWHM at science wavelengths.  For OSIRIS observations of NGC 3842 and NGC 7768, we list the FWHM values of the inner and outer PSF components.
\end{small}
\vspace{0.15in}
\end{table*}

\subsection{Photometry}
\label{sec:phot}

We measure the stellar light profiles of each BCG with a combination of $R$-band (0.6 $\mu$m) and $I$-band (0.8 $\mu$m) photometry.
For radii, $r$, out to $10''$ we adopt high-resolution surface brightness profiles from WFPC2 on the \textit{Hubble Space Telescope} (\textit{HST}) \citep{Laine}.  
At larger radii we use $R$-band data from Lauer, Postman \& Strauss (private communication), obtained with the 2.1-m telescope at Kitt Peak National Observatory (KPNO).  The KPNO data have a field-of-view (FOV) of $5.2' \times 5.2'$, which enables accurate sky subtraction.  Our ground-based surface brightness profile of each galaxy extends to $115''$.
We have combined the individual profiles from WFPC2 and KPNO at overlapping radii between $5''$ and $10''$, varying the respective weights such that the WFPC2 data contribute $100\%$ to the combined profile at $r = 5''$ and the KPNO data contribute $100\%$ at $r = 10''$.  Before stitching the profiles, we translate the WFPC2 profile to $R$-band using the average $R-I$ color between $5''$ and $10''$.  
NGC 7768 has a dust disk at $r < 0.5''$ \citep{Grillmair94}, so we
measure the innermost stellar profile with additional $H$-band (1.6 $\mu$m)
photometry from \textit{HST}/NICMOS (1997; PI Tonry).

We convert surface brightness to stellar luminosity density using the deprojection procedure of \citet{Geb96}.  In order to model each galaxy, we must define a symmetry axis.  The symmetry axis in three dimensions is projected to the minor axis on the sky, and the model galaxy's equatorial plane corresponds to the photometric major axis.  Isophotal twists are evident in the outer part of NGC 3842 and the inner part of NGC 2832; in each case, we choose the major axis 
position angle that matches the largest range of radii covered by our kinematic data.  
The major axis of NGC 4889 is near $80^\circ$ east of north at all radii.  We summarize our photometric measurements of each galaxy in Table~\ref{tab:sample}.  The major- and minor-axis luminosity density profiles are shown in Figure~\ref{fig:lden}.

We use $R$-band photometry in our stellar orbit models and therefore constrain $\mlr$.  It is useful to derive $\mlv$ so we can compare our results with other studies of mass-to-light ratios in early-type galaxies.  
To convert from $R$- to $V$-band, we use galaxy colors from \citet{PL95} and the Sloan Digital Sky Survey, with filter translations from \citet{BR07}.

\subsection{Integral-field Spectroscopy}
\label{sec:spec}

We measured kinematics in NGC 3842, NGC 4889, NGC 7768, and NGC 2832 using three different integral-field spectrographs (IFSs) to cover two orders of magnitude in radius.  Data from GMOS-North \citep{ASmith02,Hook04} on the 8-m Gemini Observatory North telescope provide the most reliable measurements of the central stellar kinematics in each BCG.  Additional observations with OSIRIS \citep{Larkin} and the laser guide star adaptive optics (LGS-AO) system \citep{vanDam,Wiz} on the 10-m W. M. Keck II telescope provide higher spatial resolution but substantially worse signal-to-noise.  We measured wide-field kinematics with the George and Cynthia Mitchell Spectrograph \citep[formerly VIRUS-P;][]{Hill} on the 2.7-m Harlan J. Smith Telescope at McDonald Observatory.  Our spectroscopic observations are summarized in Table~\ref{tab:spec}.

\subsubsection{GMOS}
\label{sec:GMOS}

\begin{figure}[!b]
\vspace{-0.1in}
  \epsfig{figure=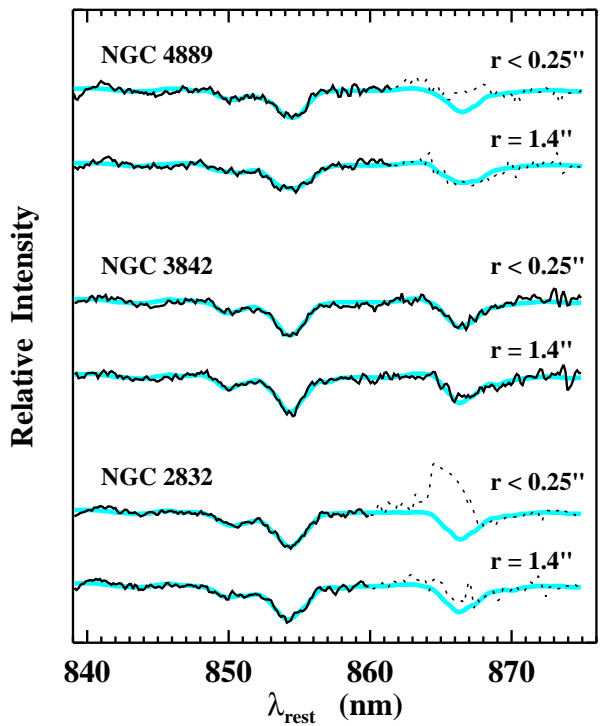,width=3.5in}
 \caption{Example spectra of the calcium triplet region from the GMOS IFS.  For each galaxy, the central spectrum ($r < 0.25''$) is displayed, as well as spectrum near the major axis, corresponding to an inner radius of $1.1''$ and an outer radius of $1.7''$.  The major-axis spectra correspond to the east side of NGC 4889, the north side of NGC 3842, and the north side of NGC 2832.  The thick cyan line for each spectrum represents the best-fitting, LOSVD-convolved stellar template.  Dotted lines indicate spectral channels that were masked during fitting.  Each spectrum was continuum-divided before fitting and plotting.}
\label{fig:Gsample}
\end{figure}

GMOS-N is a multi-purpose spectrograph on Gemini North.  We used GMOS in IFS mode, which  maps the science field and a simultaneous sky field with hexagonal lenslets.  We observed the center of each BCG with the \textit{CaT} filter, centered near the infrared CaII triplet.  A representative spectrum from each galaxy is shown in Figure~\ref{fig:Gsample}.  The GMOS data were reduced using version 1.4 of the Gemini IRAF software package\footnote{available from Gemini Observatory, at http://www.gemini.edu/sciops/data-and-results/processing-software}.  

GMOS is seeing-limited, with a lenslet scale of $0.2''$.  We measure full widths at half-maximum (FWHM) of $0.4 - 0.7''$ for point sources in acquisition frames from the GMOS imager.  For all stellar orbit models discussed below, we have assumed seeing of $0.4''$ FWHM for GMOS data.  In a preliminary trial with NGC 3842, we obtained consistent values of $\mbh$ and $\ml$ from stellar orbit models assuming $0.4''$ and $0.7''$ seeing.

%

\subsubsection{OSIRIS}
\label{sec:OSIRIS}

\begin{figure}[!b]
\vspace{-0.1in}
  \epsfig{figure=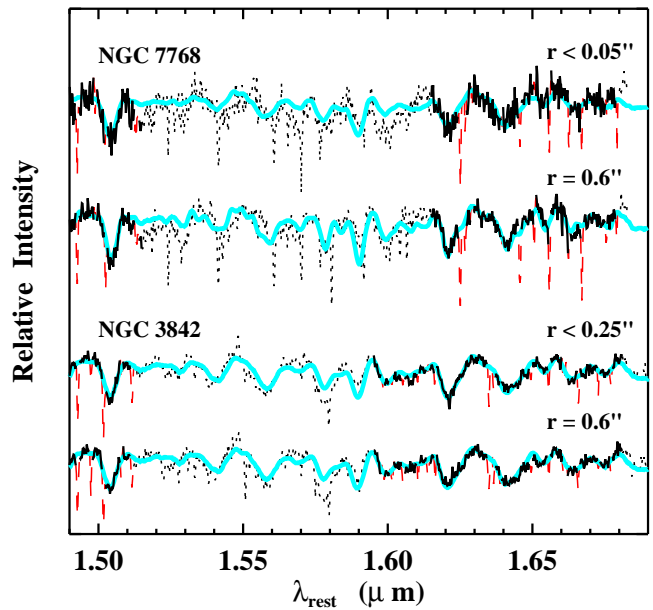,width=3.5in}
 \caption{Example $H$-band spectra from OSIRIS.  The central spectrum for NGC 7768 covers $0.1'' \times 0.1''$.  The central spectrum for NGC 3842 corresponds to $r < 0.25''$.  We also display an off-center spectrum for each galaxy; the corresponding spatial region is near the major axis, with an inner radius of $0.4''$ and an outer radius of $0.7''$.  The major-axis spectra correspond to the east side of NGC 7768 and the north side of NGC 3842.  The thick cyan line for each spectrum represents the best-fitting, LOSVD-convolved stellar template.  Dotted lines indicate spectral channels that were masked during fitting, and dashed red lines indicate residual features from telluric OH, which were also masked.  Each spectrum was continuum-divided before fitting and plotting.  OSIRIS spectra for NGC 4889 were compromised by warm detector temperatures in 2009.}
\label{fig:Osample}
\end{figure}

OSIRIS is a near-infrared (NIR), IFS built for use with the Keck AO system.  We observed each BCG with the $0.05''$ lenslet scale and broad $H$-band filter, which covers several metal absorption lines and CO and OH vibrational bandheads at observed wavelengths of 1.47-1.80 $\mu$m.  Figure~\ref{fig:Osample} shows representative OSIRIS spectra.
OSIRIS has no sky lenslets, and extended objects such as BCGs cover the entire science field.  We observed each BCG with a repeated ``object-sky-object,'' dither sequence, such that every 900-s science frame was immediately preceded or followed by a 900-s sky frame.  
We observed kinematic template stars using the same filter and lenslet scale as the BCGs.  Spatial variations in instrumental resolution are negligible relative to the velocity broadening in BCGs.

%
%

%
\begin{figure*}[!t]
  \epsfig{figure=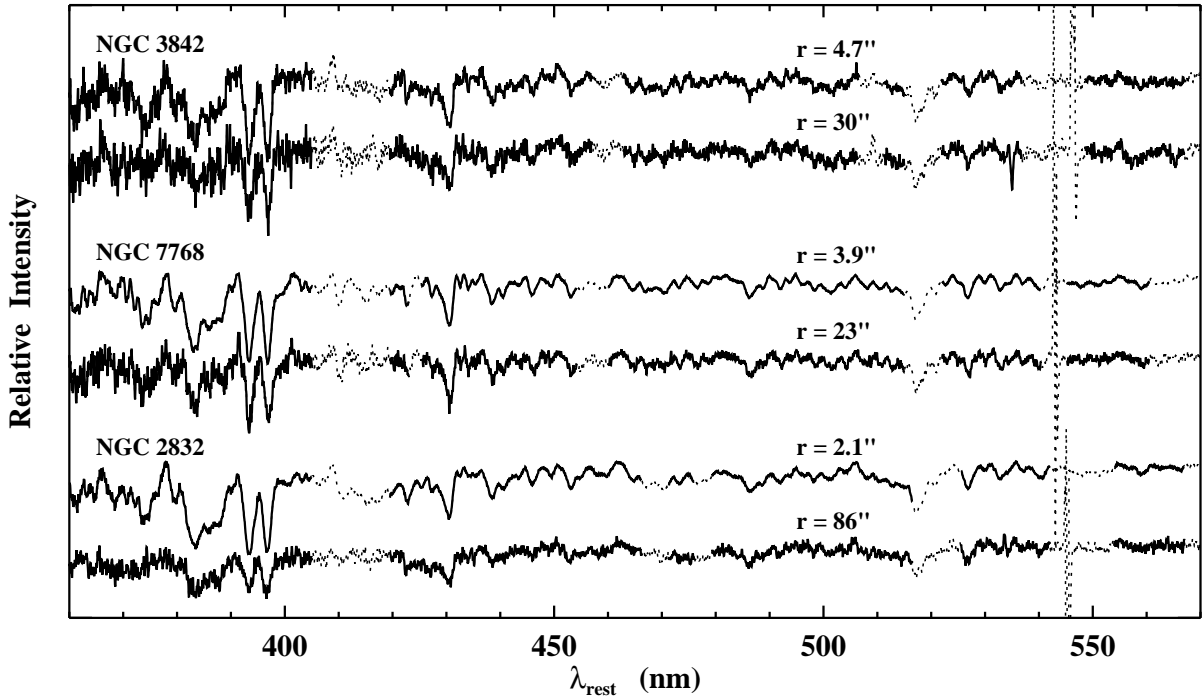,width=7.0in}
 \caption{Example spectra from the Mitchell Spectrograph.  For each galaxy, we display spectra corresponding to the smallest and largest radii where we employ Mitchell Spectrograph data in our stellar orbit models.  The corresponding bins span $3.8-5.6''$ and $24.5-35.3''$ for NGC 3842, $3.0-4.8''$ and $18.3-28.5''$ for NGC 7768, and $1.7-2.5''$ and $70-101''$ for NGC 2832.  Dotted lines indicate spectral channels that were masked during fitting.  Each spectrum was continuum-divided before fitting and plotting.}
\label{fig:Vsample}
\vspace{0.15in}
\end{figure*}

We used version 2.3 of the OSIRIS data reduction pipeline\footnote{available from the UCLA Infrared Laboratory, at http://irlab.astro.ucla.edu/osiris/pipeline.html}
to perform sky subtraction, spatial flat-fielding, spectral extraction, wavelength calibration, and spatial mosaicking of 3-dimensional data cubes.  
We used custom routines to remove bad pixels and cosmic rays from 900-s exposures and to calibrate for telluric absorption.  
We estimated the AO PSF by observing the LGS-AO tip/tilt star with the OSIRIS Imager.  For NGC 3842 and NGC 4889, we typically measured the PSF once per half-night.  For NGC 7768 we measured the PSF four times over an observing span of eight hours.  The weighted average PSF for NGC 7768 has a Strehl ratio of $22\%$ and is well-fit by an inner component with $0.04''$ FWHM and an outer component with $0.09''$ FWHM.  In a similar investigation of the BCG NGC 6086, \citet{mcconnell11a} varied the AO PSF and obtained a systematic error of only $\sim 10\%$ in the final measurement of $\mbh$.  In their investigation of M87, \citet{Geb11} measured similar values of $\mbh$ for PSFs with different Strehl ratios.

OSIRIS data for NGC 4889 and NGC 3842 have lower quality than we initially expected.  Even spectra with relatively high signal-to-noise ($S/N$) yield unconvincing kinematic measurements.  One severe noise source is telluric OH emission, which varies faster than the 900-s exposure time and leaves large residuals even after subtracting sky frames.  Although we mask the narrow residual features when measuring stellar kinematics, the missing spectral channels conceal information about the overall shapes of absorption features.  Further difficulties arise from defining a continuum level among the blend of metallic and molecular absorption features in $H$-band.  Finally, OSIRIS suffered from elevated detector temperatures throughout 2009, including the bulk of our observations of NGC 4889.  Data from this period exhibit increased dark noise and possible errors in spectral extraction and wavelength calibration.  

Our OSIRIS data for NGC 4889 do not have sufficient quality to include in stellar orbit models.  We have run stellar orbit models of NGC 3842 with and without OSIRIS data; we compare results from these models in Section~\ref{sec:BH3842}.  OSIRIS data were included in all models of NGC 7768.

\subsubsection{Mitchell Spectrograph}
\label{sec:VIRUSP}

The Mitchell Spectrograph, formerly named VIRUS-P, is an optical IFS on the 2.7-m telescope at McDonald Observatory.  It features 246 fibers spaced evenly across a $107'' \times 107''$ field of view, with a one-third filling factor.  Each fiber has a diameter of $4.1''$; this is the limiting factor in spatial resolution, rather than optical seeing.  

We observed each galaxy with the low-resolution grism, which provides wavelength coverage of $\approx$ 360-580 nm, including the Ca H + K region, the $G$-band region, H$\beta$, the Mg $b$ region, and several Fe absorption features.  The spectral resolution varies over different fibers and also with wavelength.  Our observations using the low-resolution grism span instrumental resolution values of $\approx 0.45-0.65$ nm FWHM, with a corresponding range of $\sigma \sim 100-170 \kms$.  This is sufficient to resolve the velocity profiles of BCGs ($\sigma \sim 250 - 350 \kms$).  Because we employed the Mitchell Spectrograph for wide-field coverage rather than fine spatial resolution, we did not perform any sub-dithers to fill the gaps between fibers.

We used the Vaccine data reduction pipeline to perform bias subtraction and flat-fielding, compute the wavelength solution for each fiber, extract a spectrum for each fiber, model and subtract the sky spectrum, and reject cosmic rays.  Detailed descriptions of the data reduction procedures can be found in \citet{Adams11} and \citet{Murphy11}.

The maximum radius at which Mitchell Spectrograph data can yield robust kinematic measurements depends on the surface brightness profile of the galaxy, the background sky conditions, and the overall integration time at a given pointing.  For each galaxy, we have binned numerous fibers at large radii and preserved data for which a good kinematic fit was recoverable for at least one binning scheme.
Following this procedure, our outermost bins for kinematic extraction cover radii of 
$24.5-35.3''$ for NGC 3842,
$18.3-28.5''$ for NGC 7768, and
$70-101''$ for NGC 2832.
Sample spectra from the central and outer regions of each galaxy are shown in Figure~\ref{fig:Vsample}.

%
\section{Extracting Stellar Kinematics}
\label{sec:extract}

Our dynamical models fit weighted and superposed stellar orbits to line-of-sight velocity distributions (LOSVDs) extracted from spectroscopic data.  We extract LOSVDs with a Maximum Penalized Likelihood (MPL) technique, which fits an LOSVD-convolved stellar template to each galaxy spectrum.  The LOSVDs are non-parametric, defined at 15 radial velocity bins in our orbit models.  The MPL fitting method is described in detail in \citet{Geb00b}, \citet{Pinkney}, and \citet{Nowak08}, and adaptations for integral field data of BCGs are described in \citet{mcconnell11a}.  

In order to attain sufficient signal-to-noise ($S/N$) for effective kinematic extraction, we perform spatial binning on each data set.  For GMOS data of NGC 4889, NGC 3842 and NGC 2832, our central bin combines 7 hexagonal lenslets; the corresponding bin diameter is $0.55''$.  For OSIRIS data of NGC 7768, we combine 4 square lenslets to obtain a central bin of $0.1'' \times 0.1''$.  At large radii (typically $r > 15''$) we combine Mitchell Spectrograph data from multiple angular bins. 

A spectral binning factor is necessary to smooth over channel-to-channel noise in some spectra.  We typically use smoothing factors of 2 spectral pixels for GMOS data and 8-10 spectral pixels for OSIRIS data.  Lower smoothing factors yield jagged LOSVDs.  Mitchell Spectrograph spectra do not require smoothing.  

The near-infrared CaII triplet has a clearly defined continuum and yields robust kinematic measurements across a broad range of stellar templates \citep{Dressler84, Barth02}.  Therefore, we fit GMOS spectra with a single G9III template star.  In spectra of NGC 4889 and NGC 2832, the 866 nm calcium line is contaminated by a detector artifact and is not included our fits (see Fig.~\ref{fig:Gsample}).  The 850 nm and 854 nm lines are included in all fits.  

For OSIRIS spectra, we optimize our LOSVD extraction by carefully choosing a subset of the many  absorption features in $H$-band.  Although we are able to perform a global equivalent width adjustment before fitting stellar templates, some of the individual metal lines exhibit further mismatch between the galaxy and template equivalent widths.  Some molecular features are blended with the suspect metal lines, and others are compromised by residuals from telluric OH subtraction.  The consistently reliable features for kinematic extraction are the $\nu$ = 3-6,  $\nu$ = 4-7, and $\nu$ = 5-8 bandheads of $^{12}$CO, spanning the sub-region from 1.61 to 1.68 $\mu$m rest, and the MgI absorption feature near 1.504 $\mu$m rest (see Fig.~\ref{fig:Osample}).  We fit OSIRIS spectra using a set of nine template stars with spectral types from G8 to M4.  Template weights are allowed to vary freely and are fit simultaneously with the LOSVD.

Mitchell Spectrograph spectra contain a multitude of absorption features, which are dominated by different stellar types and require different equivalent width adjustments.  Following the procedure of \citet{Murphy11}, we divide each spectrum into five spectral sub-regions and extract LOSVDs independently for each sub-region.  Our fits use a set of 16 template stars selected from the Indo-US library \citep{Valdes}, with spectral types from B9 to M3.  We compute the instrumental resolution as a function of wavelength in each fiber and convolve the template spectra with an appropriately weighted instrumental resolution profile for each galaxy spectrum.  After fitting each galaxy spectrum, we discard sub-regions with visually unconvincing fits or severely asymmetric LOSVDs.  The LOSVDs from the remaining sub-regions are averaged to produce a final representative LOSVD.

We perform Monte-Carlo trials to determine the uncertainties in the best-fit LOSVDs.  In each trial, random noise is scaled to the root-mean-squared residual of the original fit and added to the galaxy spectrum before re-fitting.  In each bin of the LOSVD, uncertainties are computed from the distribution of trial values.  Our fitting method sometimes yields LOSVDs with noise at large positive or negative velocities.  Consequently, we force our uncertainties to extend to zero power in the LOSVD wings.  For Mitchell Spectrograph LOSVDs, we adopt the average uncertainty from the individual spectral sub-regions, or if greater, the 1-$\sigma$ scatter in the sub-regions' LOSVD values.  

\bigskip

%
\section{Two-Dimensional Kinematics}
\label{sec:kin}

For each galaxy, we fit a 4th-order Gauss-Hermite polynomial to each
non-parametric LOSVD in order to illustrate the stellar kinematics.  Below
we highlight spatial trends in the Gauss-Hermite moments $v_{\rm rad}$,
$\sigma$, $h_3$, and $h_4$.  To measure the Gauss-Hermite moments, the LOSVD is fit with a function $f(v)$, defined by
\begin{eqnarray}
  f(v) & \propto & \frac{1}{\sqrt{2\pi \sigma^2}} e^{-(v-v_{\rm rad})^2/\sigma^2}
              \\
     &\times&  \left[ 1 + h_3 H_3\left( \frac{v-v_{\rm rad}}{\sigma}\right) + 
        h_4 H_4\left( \frac{v-v_{\rm rad}}{\sigma} \right) \right] \,, \nonumber
\end{eqnarray}
where $H_3(x)=\frac{1}{\sqrt{3}}(2x^3-3x)$ and\\
$H_4(x)=\frac{1}{\sqrt{24}}(4x^4-12x^2+3)$.

\bigskip

%
\begin{figure*}[!t]
\centering
\vspace{-0.2in}
  \epsfig{figure=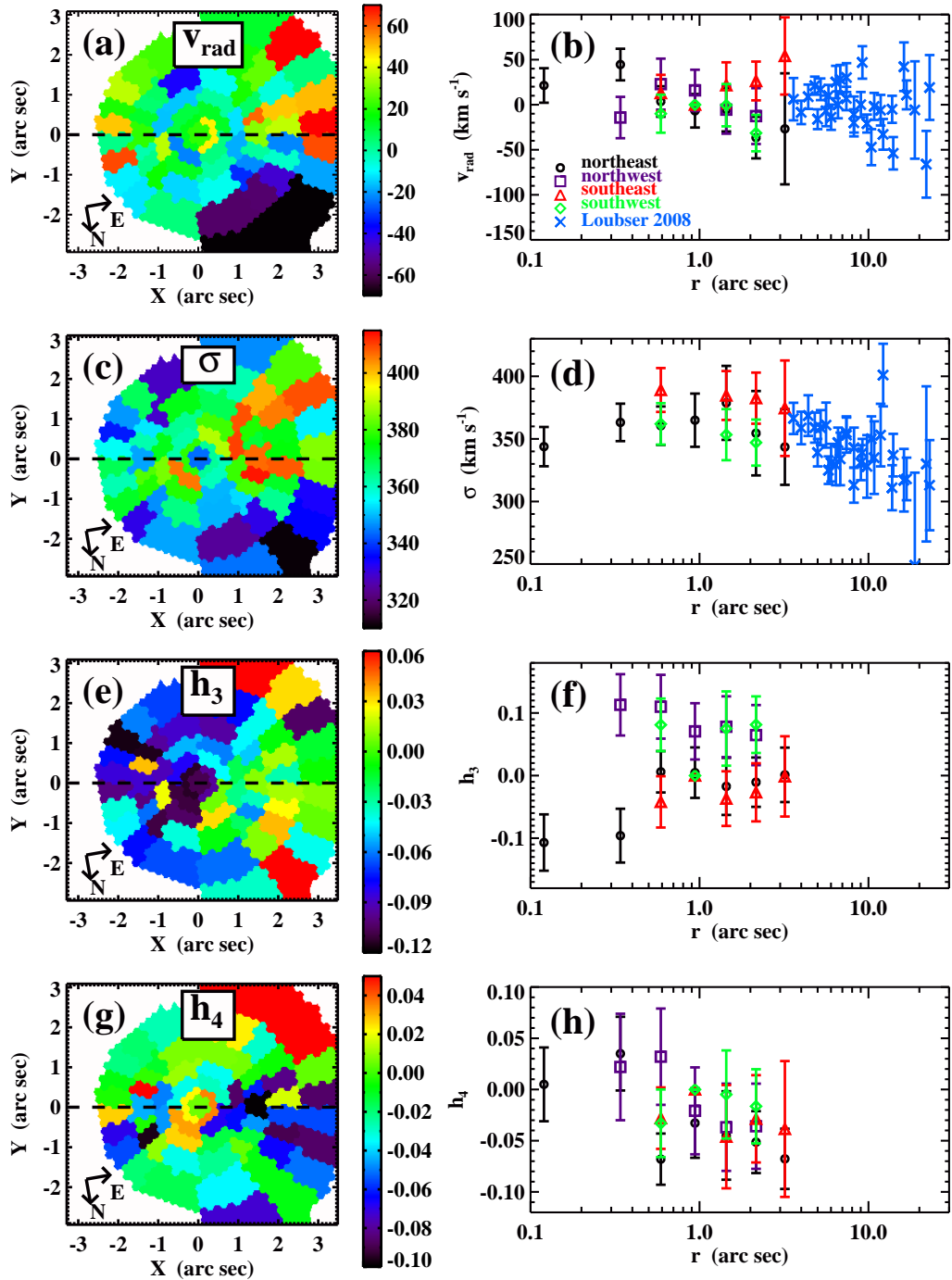,width=5.5in}
 \caption{Stellar kinematics in NGC 4889.  Panels (a), (c), (e), and (g) are 2-dimensional maps from GMOS IFS data.  The horizontal dashed line represents the photometric major axis.  Panels (b), (d), (f), and (h) show kinematic moments as a function of radius, after averaging kinematic moments from different polar angles in each quadrant of NGC 4889.   Panels (b) and (d) also include major-axis kinematic moments from \citet{Loubser}.  For the 1-dimensional plots, the values of $v_{\rm rad}$ and $h_3$ have been inverted on the west side of the galaxy.   (a) and (b): radial velocity.  (c) and (d): line-of-sight velocity dispersion.  (e) and (f):  Gauss-Hermite moment $h_3$.  (g) and (h):  Gauss-Hermite moment $h_4$.}
\label{fig:kin4889}
\vspace{0.1in}
\end{figure*}
%

%
\begin{figure}[!b]
\hspace{-0.4cm}
  \epsfig{figure=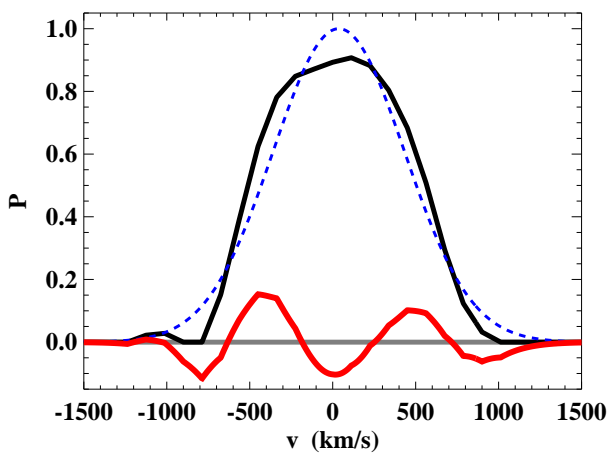,width=3.5in} 
\vspace{-0.4cm}
\caption{LOSVD near the velocity dispersion peak in NGC 4889, corresponding to a radius of $1.4"$ (700 pc) along the east side of the major axis.  The solid black line is the non-parametric LOSVD, and the dashed blue line is the best-fit Gaussian profile.  The thick red line represents the residual velocity profile after the Gaussian profile is subtracted.  The residual profile includes tangential orbits centered near $\pm 450 \kms$.  Negative features in the residual profile represent an under-abundance of stars at the corresponding line-of-sight velocities.  
}
\label{fig:los4889}
\end{figure}
%

%
\begin{figure*}[!t]
\begin{center}
  \epsfig{figure=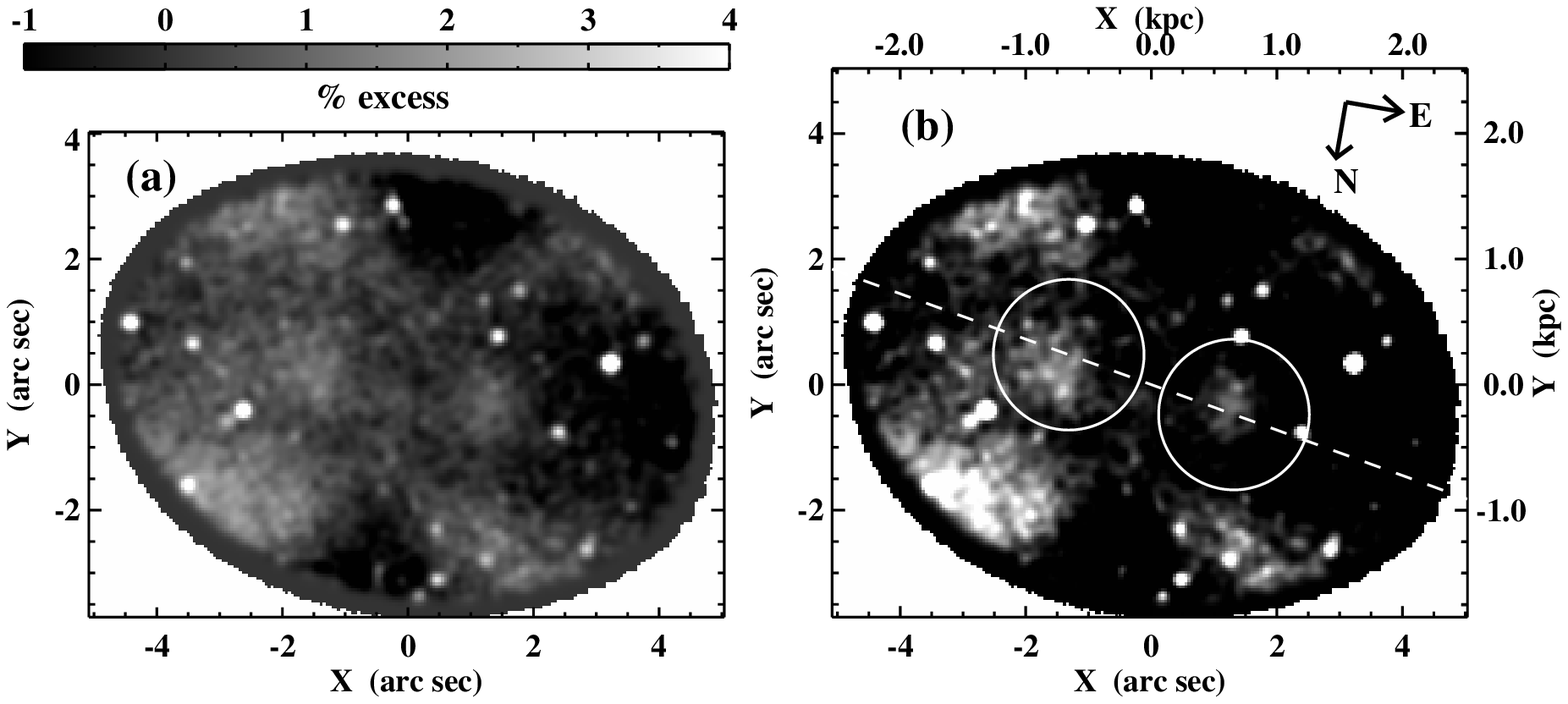,width=6.5in} 
\end{center}
\vspace{-0.7cm}
\caption{(a) Residual image of NGC 4889, after fitting elliptical isophotes to a de-convolved image from \textit{HST}/WFPC2 (F606W).  (b) Same image, with contrast adjusted to highlight the $\sim 1\%$ excess light from a possible stellar torus.  The dashed white line indicates the major axis of NGC 4889.  The white circles are each centered on the major axis $1.4''$ from the galaxy center.  Each circle has a radius of $1.2''$.  The compact bright spots in both panels are globular clusters.}
\label{fig:phot4889}
\vspace{0.1in}
\end{figure*}

\subsection{NGC 4889}
\label{sec:losvd4889}

Figure~\ref{fig:kin4889} shows two-dimensional maps of $v_{\rm rad}$, $\sigma$, $h_3$, and $h_4$, from GMOS observations of NGC 4889.  Figure~\ref{fig:kin4889} also includes radial profiles of each moment in NGC 4889.  The radial profiles include our GMOS data and measurements from \citet{Loubser}, recorded with the long-slit spectrograph ISIS at the William Herschel Telescope (WHT).  

The kinematic moments in NGC 4889 show several asymmetries with respect to the major and minor axis.  Asymmetries in the line-of-sight velocity dispersion, $\sigma$, likely present the greatest difficulties for accurately measuring $\mbh$.  On the east side of the galaxy, $\sigma$ peaks at $413 \pm 22 \kms$ near the major axis and remains above $400 \kms$ through an extended region in the southeast quadrant.  On the west side, $\sigma$ exceeds $385 \kms$ in only a single spatial bin, reaching $406 \pm 18 \kms$ at $0.6''$ toward the northwest.  At the very center of NGC 4889, we measure $\sigma = 344 \pm 16 \kms$.  
A strong deficit of radial orbits can produce the central drop in line-of-sight velocity dispersion even in the presence of a supermassive black hole; this is further discussed in Section~\ref{sec:orbits}.  Central decreases in velocity dispersion have been observed in several other early-type galaxies \citep[e.g.,][]{vdM94,Pinkney,Houghton,Geb07,Nowak08}. 

The $h_4$ moment, which describes whether an LOSVD is boxy ($h_4 < 0$) or peaky ($h_4 > 0$), varies with radius and polar angle in NGC 4889.  At radii from $0.7''$ to $3.5''$, $h_4$ is negative near the major axis and nearly zero toward the minor axis.    
We have fit and subtracted a Gaussian profile from each non-parametric LOSVD and have examined the residual velocity profiles.  In most of the spatial bins near the major axis, the residual profiles contain peaks near $\pm 450 \kms$, as exemplified in Figure~\ref{fig:los4889}.
An LOSVD sub-component with velocity peaks near $\pm 450 \kms$ could correspond to a torus or thick ring of stars, including rotating and counter-rotating populations.  In an eccentric ring, $\sigma$ would exhibit a local maximum near periapsis. 

To further investigate the hypothesis that NGC 4889 hosts a central stellar torus, we have carefully examined the inner light profile of the galaxy.  We have determined a best-fit surface brightness profile by fitting concentric ellipses to a reduced and de-convolved image from \textit{HST}/WFPC2, as in \citet{Laine}.  Subtracting the best-fit ellipses yields a 2-dimensional residual image of NGC 4889, displayed in Figure~\ref{fig:phot4889}.  The residual image shows two peaks along the major axis, which straddle the galaxy center and contain approximately 
$1\%$ of the total light at their corresponding positions.  
These peaks resemble a diffuse torus of stars in the equatorial plane of the galaxy, spanning radii from $1''$ to $2''$.
On the east side of NGC 4889, the residual photometric peak overlaps with the velocity dispersion peak but does not extend to match the extended pattern of high $\sigma$ in the southeast quadrant.  Boxy LOSVDs overlap with both residual photometric peaks but also extend to radii beyond $2''$.  Although NGC 4889 could host a stellar torus with a double-peaked velocity profile, the torus would only contribute $\sim 1\%$ of the total signal in the corresponding LOSVDs, too little to be fully responsible for the LOSVDs' non-Gaussianity.

The double-peaked feature in our residual image of NGC 4889 appears to be qualitatively similar to the central structure seen in a number of high-luminosity galaxies, such as 
NGC 4073, NGC 6876, and the BCGs NGC 910, IC 1733, and IC 4329 \citep{Lauer02, Lauer05}.
In these galaxies the surface brightness profile actually has a local minimum at the center.  \citet{Lauer02} have argued that such an appearance could be due to the addition of a diffuse torus to the cores of the galaxies, or could alternatively be evidence for the evacuation of stars from the center by a merging pair of supermassive black holes bound in a tight binary.  In the case of NCG 4889, the torus is not strong enough to create an apparent surface brightness minimum at the center of the galaxy, and would thus be a weaker example of the phenomenon seen in the \citet{Lauer02, Lauer05} galaxies.
The radius of the torus in NGC 4889 is $\sim \rinf$.  In simulations by \citet{ZB01}, a black hole binary with initial separation $\rinf$ scours a torus of radius $\sim 3 \rinf$ on a timescale comparable to the hardening and coalescence of the binary.

Alternative scenarios for asymmetric kinematics include an ongoing minor merger, a surviving nucleus from a cannibalized satellite galaxy, or a chance alignment with a foreground satellite.  However, none of these scenarios are consistent with the highly regular photometric contours in NGC 4889: even the $1\%$ deviations from perfect ellipses are distributed symmetrically about the galaxy center.  Another post-merger scenario is a black hole displaced from the center by gravitational wave recoil.
Simulations of merger remnants with recoiling black holes suggest that even with a modest recoil velocity, a black hole can maintain an observable displacement in a gas-poor galaxy \citep[e.g.,][]{Blecha11, Guedes11, Sijacki11}, and damping could be particularly weak with a shallow stellar mass profile like the core of NGC 4889.  However, if a recoiling black hole affected a sufficient number of stars to displace the galaxy's velocity dispersion peak, we would expect to observe a similarly prominent photometric disturbance, or even a cusp of stars surrounding the black hole.   

Although we do not find an obvious explanation for our asymmetric velocity dispersion measurements in NGC 4889, we suggest that they arise from an asymmetric orbital structure within a largely symmetric spatial distribution of stars.  In Section~\ref{sec:orbits}, we illustrate how tangential orbital bias in the vicinity of the black hole can produce a central minimum in the line-of-sight velocity dispersion.  A stellar torus is consistent with excess tangential orbits as well as our photometric data.  However, a circular or slightly eccentric torus of stars cannot fully explain the velocity dispersion features in the southeast and northwest quadrants of NGC 4889 (see Fig.~\ref{fig:kin4889}c).

%
\begin{figure*}[!t]
\centering
\vspace{-0.2in}
  \epsfig{figure=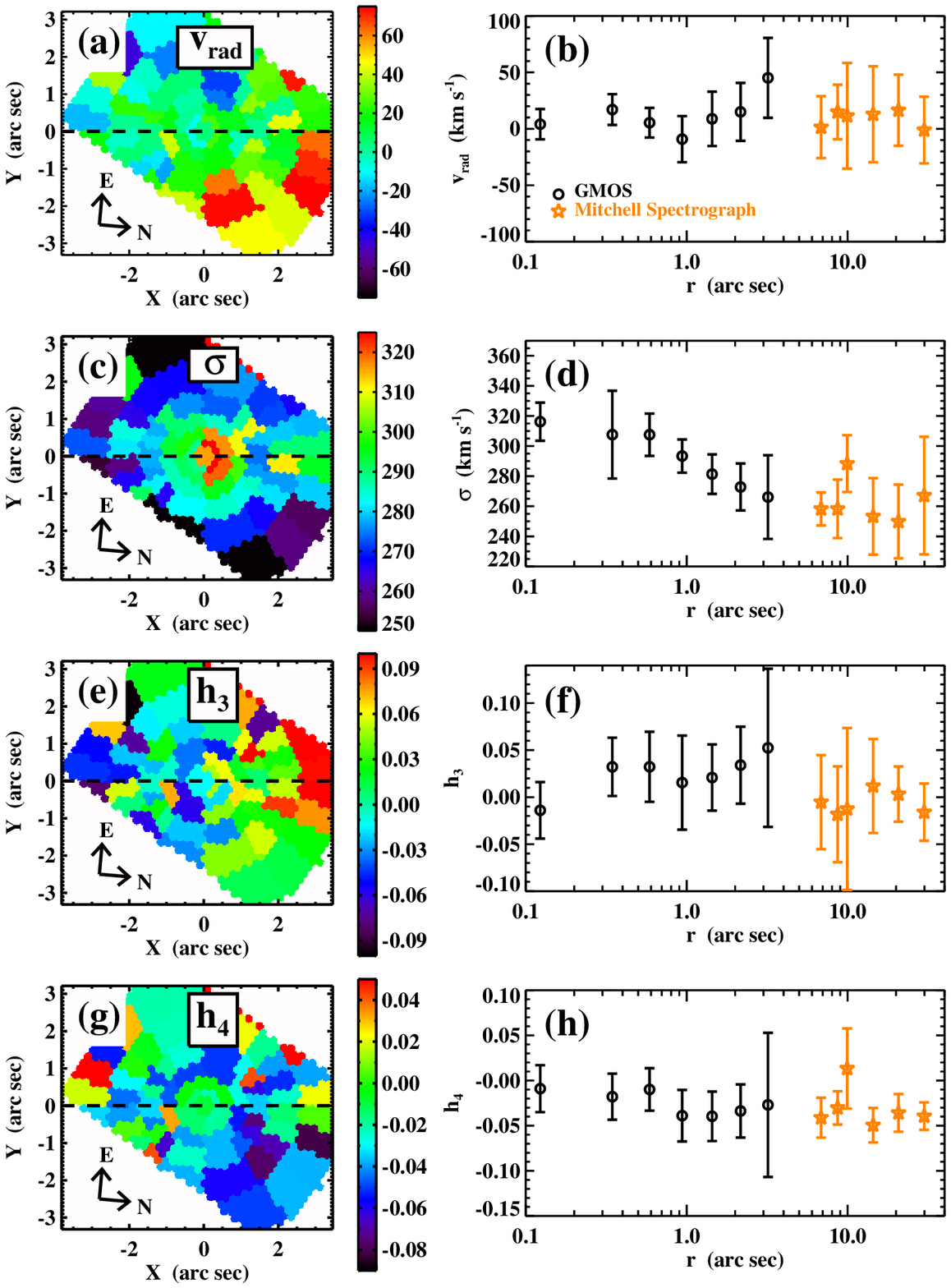,width=5.5in}
 \caption{Stellar kinematics in NGC 3842.  Panels (a), (c), (e), and (g) are 2-dimensional maps from GMOS IFS data.  The horizontal dashed line represents the photometric major axis.  Panels (b), (d), (f), and (h) are radial plots, including GMOS and Mitchell Spectrograph data.  In order to construct the radial plots, the kinematic moments from different quadrants and polar angles have been averaged, after inverting the values of $v_{\rm rad}$ and $h_3$ in the southeast and southwest quadrants.   (a) and (b): radial velocity.  (c) and (d): line-of-sight velocity dispersion.  (e) and (f):  Gauss-Hermite moment $h_3$.  (g) and (h):  Gauss-Hermite moment $h_4$.}
\label{fig:kin3842}
\vspace{0.1in}
\end{figure*}
%

%
\begin{figure*}[!t]
\centering
\vspace{-0.2in}
    \epsfig{figure=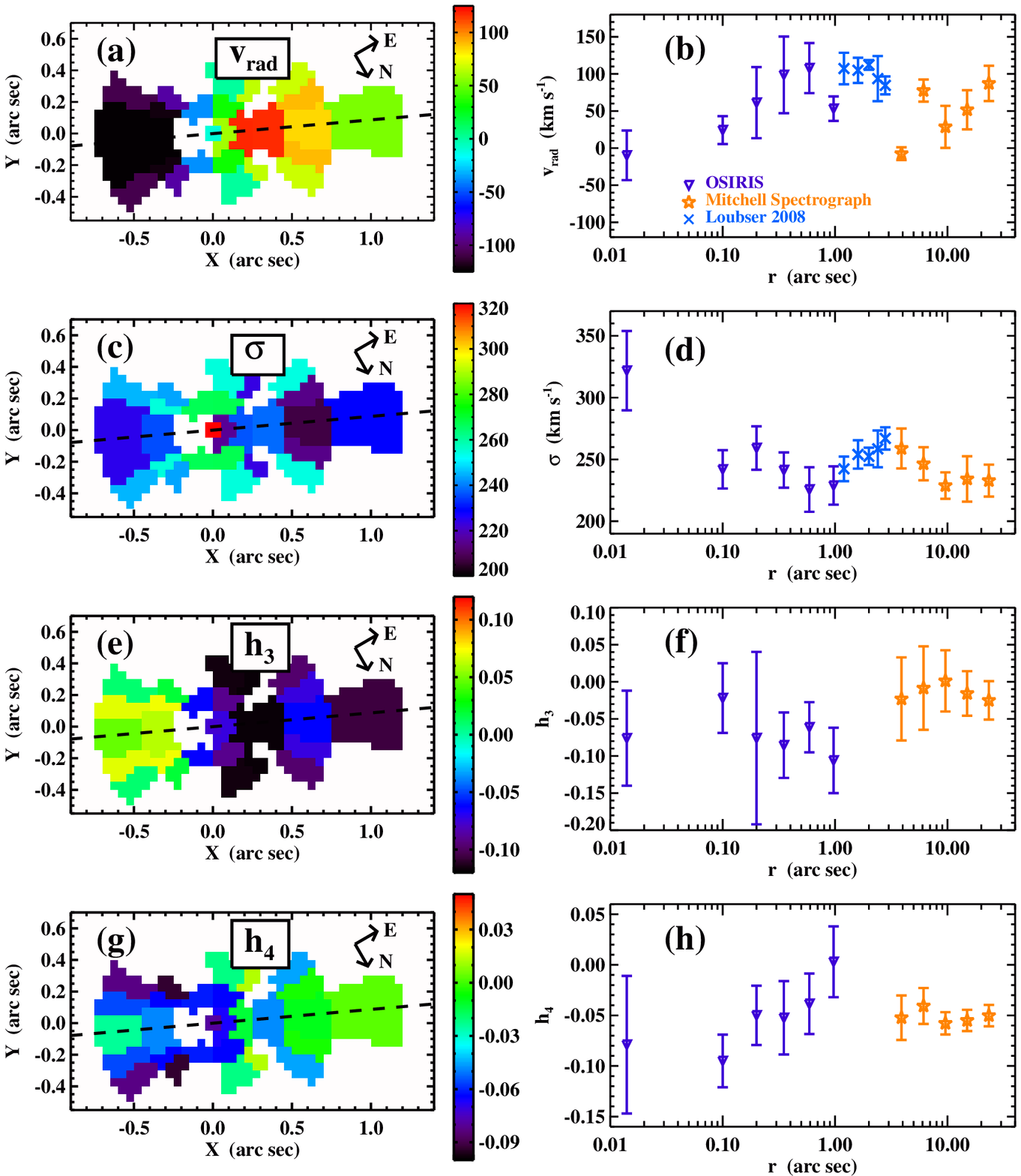,width=6.4in}
 \caption{Stellar kinematics in NGC 7768.  Panels (a), (c), (e), and (g) are 2-dimensional maps from OSIRIS IFS data.  The horizontal dashed line represents the photometric major axis.  Panels (b), (d), (f), and (h) are radial plots, including OSIRIS and Mitchell Spectrograph data, plus long-slit data from \citet{Loubser} at radii from $1''$ to $3''$.  In order to construct the radial plots, the kinematic moments from different hemispheres and polar angles have been averaged, after inverting the values of $v_{\rm rad}$ and $h_3$ on the west side of the galaxy.   (a) and (b): radial velocity.  (c) and (d): line-of-sight velocity dispersion.  (e) and (f):  Gauss-Hermite moment $h_3$.  (g) and (h):  Gauss-Hermite moment $h_4$.}
\label{fig:kin7768}
\vspace{0.1in}
\end{figure*}
%

%
\begin{figure*}[!t]
\centering
\vspace{-0.2in}
    \epsfig{figure=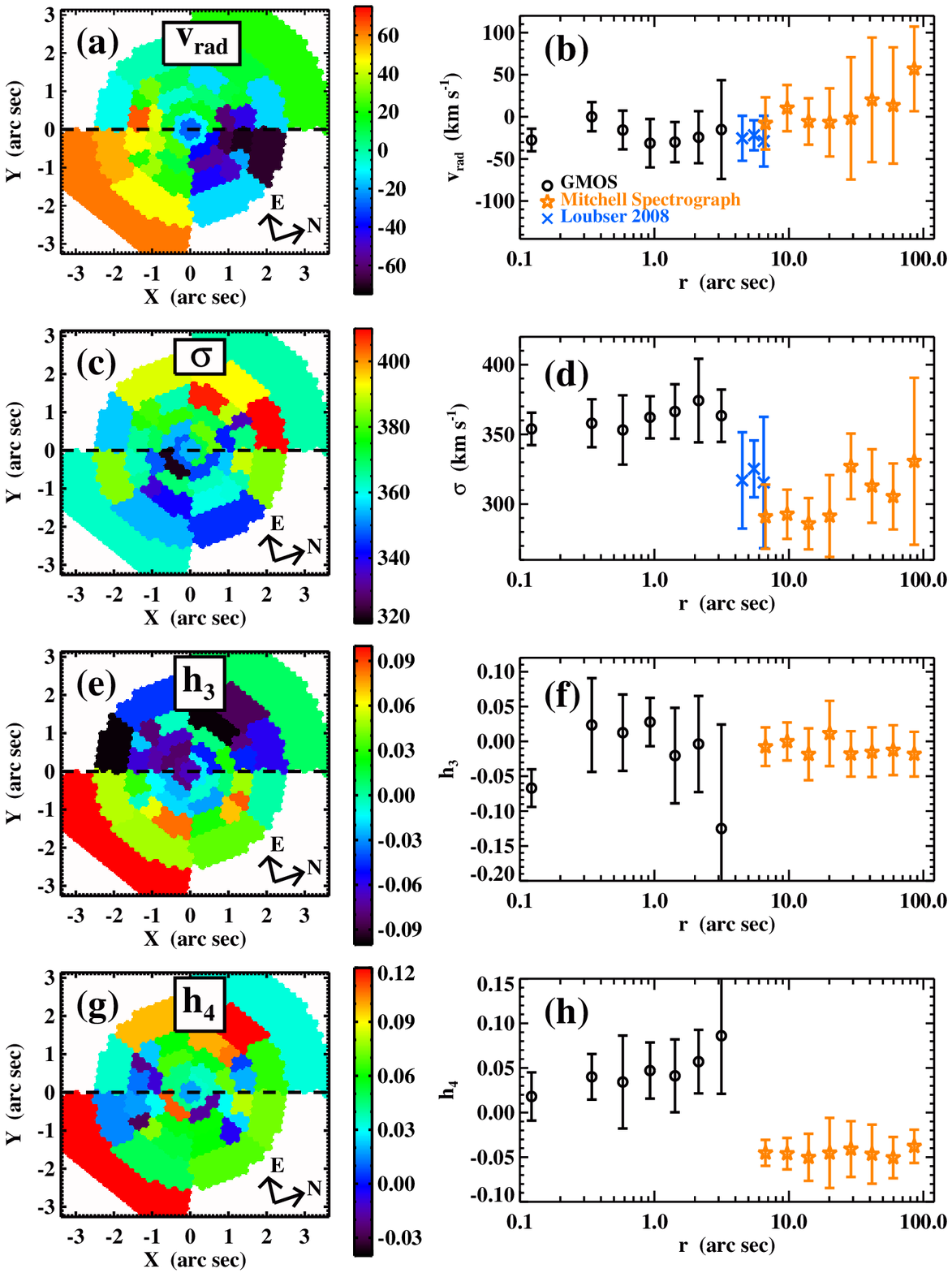,width=5.5in}
 \caption{Stellar kinematics in NGC 2832.  Panels (a), (c), (e), and (g) are 2-dimensional maps from GMOS IFS data.  The horizontal dashed line represents the photometric major axis.  Panels (b), (d), (f), and (h) are radial plots, including GMOS and Mitchell Spectrograph data, plus long-slit data from \citet{Loubser} at radii from $4''$ to $7''$.  In order to construct the radial plots, the kinematic moments from different quadrants and polar angles have been averaged, after inverting the values of $v_{\rm rad}$ and $h_3$ in the southeast and southwest quadrants.   (a) and (b): radial velocity.  (c) and (d): line-of-sight velocity dispersion.  (e) and (f):  Gauss-Hermite moment $h_3$.  (g) and (h):  Gauss-Hermite moment $h_4$.}
\label{fig:kin2832}
\vspace{0.1in}
\end{figure*}
%

\subsection{NGC 3842}
\label{sec:losvd3842}

Figure~\ref{fig:kin3842} shows two-dimensional maps of the kinematic moments in NGC 3842, as measured by GMOS, as well as radial profiles of each moment as measured by GMOS and the Mitchell Spectrograph.  NGC 3842 exhibits the simplest kinematics of the four BCGs.  There is negligible rotation, and $\sigma$ increases virtually monotonically toward the center.  
The maps of $h_3$ and $h_4$ are largely featureless, and $h_4$ is slightly negative on average.  The four quadrants of NGC 3842 are similar enough to average their kinematics and model a single set of LOSVDs.  The kinematics from the Mitchell Spectrograph are consistent with a continuation of the trends observed at small radii with GMOS.

\subsection{NGC 7768}
\label{sec:losvd7768}

Figure~\ref{fig:kin7768} shows two-dimensional maps of the kinematic moments from OSIRIS observations of NGC 7768, and radial profiles of each moment from OSIRIS and the Mitchell Spectrograph.  Unlike most BCGs, NGC 7768 exhibits a strong rotation curve along the major axis, with $|v_{\rm rad}| \sim 100 \kms$ at radii beyond $0.2''$.  We measure $\sigma = 322 \pm 32 \kms$ in our central OSIRIS bin, outside of which $\sigma$ drops suddenly and remains near 230-260$\kms$ over the full extent of our radial coverage.  We measure $h_3$ to be anti-correlated with $v_{\rm rad}$, particularly for OSIRIS data.  This anti-correlation is common in rotating early-type galaxies \citep[e.g.,][]{BSG94,Krajnovic11}.  OSIRIS and the Mitchell Spectrograph both measure boxy LOSVDs ($h_4 < 0$).

\citet{Loubser} observed NGC 7768 with ISIS on WHT, measuring major-axis kinematics out to $11.4''$.  At radii of 0.1-$1''$, we measure a steeper radial velocity gradient and lower velocity dispersion values with OSIRIS.  The differences between our measurements and those of \citet{Loubser} are reconciled by considering the respective spatial resolutions of the OSIRIS and WHT observations: whereas OSIRIS is nearly diffraction-limited, \citet{Loubser} reported $1''$ seeing at WHT.  Although the poorly resolved velocity gradient artificially increases the WHT velocity dispersion measurements, the quadratic sum $v^2_{\rm rad} + \sigma^2$ is consistent for OSIRIS and WHT data.  Only OSIRIS detects the central rise to $\sigma > 300 \kms$.

Similarly, we attribute the low value of $v_{\rm rad}$ in our innermost Mitchell Spectrograph bin (Figure~\ref{fig:kin7768}b) to an overlap between the corresponding fiber and the central velocity dispersion gradient.  We have included this data point in stellar orbit models of NGC 7768, along with five measurements from \citet{Loubser} at radii between $1''$ and $3''$.  This combination of data is illustrated in the radial plots in Figure~\ref{fig:kin7768}.  In a separate trial, we replaced the innermost LOSVD from the Mitchell Spectrograph with three additional measurements from \cite{Loubser}, covering radii of 3.0-$5.1''$.  All trials yielded consistent values of $\mbh$ and $\mlr$.

\subsection{NGC 2832}
\label{sec:losvd2832}

Figure~\ref{fig:kin2832} shows two-dimensional maps of the kinematic moments from GMOS observations of NGC 2832, as well as radial profiles from GMOS and the Mitchell Spectrograph. 
Unlike NGC 4889 and NGC 3842, which exhibit either a central or off-center local maximum in $\sigma$, our velocity dispersion map of NGC 2832 is nearly featureless, with values near $360 \kms$ out to $r \approx 3''$.  
GMOS IFS data for NGC 2832 show some evidence of rotation from north to south, although this signal is strangely absent on the east side of the galaxy.  There is a similar discrepancy in our measurements of $h_3$: we measure $h_3 < 0$ on the east side and $h_3 > 0$ on the west side.

\citet{Loubser} observed NGC 2832 with GMOS in long-slit mode, with the slit oriented $67^\circ$ from the galaxy's major axis.  They detected a kinematically decoupled core, characterized by rotation out to approximately $4''$ along the slit axis.  Our measurements of $v_{\rm rad}$ agree with the long-slit data on the west side of the galaxy, but do not follow the same trend on the east side.  IFS and long-slit measurements both detect a flat velocity dispersion profile for $r \leq 3''$, with similar average values of $\sigma$. 
We measure significantly lower velocity dispersions with the Mitchell Spectrograph, in the range $\sim 260-320 \kms$.  

The majority of our Mitchell Spectrograph measurements are at $r \geq 7''$ and do not overlap with the GMOS IFS field of view.  We use the long-slit data from \citet{Loubser} to help bridge the apparent gap between the GMOS and Mitchell Spectrograph IFS data.  For radii between $4''$ and $7''$ we have averaged multiple long-slit data points from \citet{Loubser} and produced three LOSVDs, which we include in our stellar orbit models of NGC 2832.  The corresponding values of $v_{\rm rad}$ and $\sigma$ are included in Figures~\ref{fig:kin2832}b and~\ref{fig:kin2832}d.   

Although measurements by \citet{Loubser} indicate that $\sigma$ may indeed fall from $\sim 360 \kms$ near the center of NGC 2832 to $\sim 280 \kms$ at $r \sim 10''$,
additional discrepancies point to possible systematic errors in our kinematics from GMOS and/or the Mitchell Spectrograph.  Our LOSVDs from GMOS are peaky ($h_4 > 0$), whereas our LOSVDs from the Mitchell Spectrograph are boxy ($h_4 < 0$).  Furthermore, the two Mitchell Spectrograph fibers nearest the galaxy center yield significantly lower velocity dispersions than GMOS IFS or long-slit data at overlapping radii.  For all galaxies, we exclude the innermost Mitchell Spectrograph data from our models, so as not to dilute the more highly resolved GMOS data.

Stellar template mismatch or errant treatment of the galaxy and stellar template spectra can bias both $\sigma$ and $h_4$, \citep[e.g.,][]{RW92,vdMF93,Carter,Emsellem}.  Although we have attempted to treat GMOS IFS and Mitchell Spectrograph data consistently for each of the three BCGs, NGC 2832 exhibits a particularly large number of internal inconsistencies.  This is considered in our final assessment of stellar orbit models for NGC 2832 (Section~\ref{sec:BH2832}).

%
\section{Stellar Orbit Models and Black Hole Masses}
\label{sec:results}

\subsection{Axisymmetric Orbit Models}
\label{sec:model}

We generate stellar orbit models of each galaxy using the static potential method introduced by
\citet{Schild}.  We use the axisymmetric modeling algorithm described in detail in
Gebhardt et al. (2000b; 2003), Thomas et al. (2004; 2005), and \citet{Siopis}.  Similar models are presented in \citet{RT84}, \citet{Rix97}, \citet{vdM98}, \citet{Cretton}, and \citet{VME04}.

We assume that each galaxy includes three mass
components -- stars, a central black hole, and an extended dark matter halo
-- described by the radial density profile
\begin{equation}
  \rho(r) = \frac{M_\star}{L_R} \nu(r) +  M_\bullet \delta(r) + \rho_{\rm halo}(r) \, ,
\label{eq:rho}
\end{equation}
where $\nu(r)$ is the observed luminosity density (see Figure~\ref{fig:lden}).
The resulting gravitational potential depends on $\mbh$, the stellar mass-to-light ratio $\ml$ ($\mlr$ for $R$-band photometry), and two parameters describing the dark matter density profile $\rho_{\rm halo}(r)$.  Our models are symmetric about the $z$-axis (corresponding to the projected minor axis) and the equatorial plane ($z = 0$).
Stellar orbits are generated by propagating test particles through the potential and computing their time-averaged velocities in each bin in a polar grid.  Typical models produce $\sim 30,000$ bound orbits, including both signs of the angular momentum component $L_z$.
 
To match each observed LOSVD, we compute a model LOSVD from the projected velocities of individual orbits in the spatial region corresponding to the extracted spectrum and assumed PSF.   
Each orbit is assigned a scalar weight, and the orbital weights are varied to optimize the fit between the  observed and model LOSVDs.  The goodness of fit for each model is characterized by the $\chi^2$ parameter:
\begin{equation} \chi^2 = \sum_i^{N_b} \sum_j \frac{\left[ \mathcal{L}_{i, \rm data}\left( v_j \right) - \mathcal{L}_{i, \rm model}\left( v_j \right) \right] ^2 }{\sigma^2_i \left( v_j \right)}  \, ,
\label{eq:chi2}
\end{equation}
where $\mathcal{L}_{i,\rm data}$ and $\mathcal{L}_{i,\rm model}$ are LOSVDs in each of the $i = 1, \, ... \, N_b$ spatial bins, and $\sigma^2_i$($v_j$) is the squared uncertainty in $\mathcal{L}_{i,\rm data}$ at velocity bin $v_j$.  The weights are constrained such that the summed spatial distribution of all weighted orbits must match the observed luminosity density profile.   

For each galaxy, we have run multiple trials of orbit models.  Each trial
assumes a particular dark matter halo profile and fits a specific
combination of kinematic data to several hundred models covering a finely
sampled grid in $\mbh$ and $\ml$.  The resulting confidence limits in
$\mbh$ and $\ml$ are computed by analyzing the distribution of $\chi^2$
values, according to the cumulative likelihood method of
\citet{mcconnell11a}.  Our modeling trials for the four galaxies are
  summarized in Tables~\ref{tab:res4889}-\ref{tab:res2832}.

High-resolution and wide-field data play complementary roles in constraining the gravitational potential of each BCG.  Kinematic measurements from GMOS are sufficient to detect a black hole in NGC 4889 and NGC 3842, in part because of excellent seeing.  Still, the black hole's gravitational signature is not fully decoupled from the enclosed stellar mass, and improved measurements of $\ml$ yield tighter confidence intervals in $\mbh$.  
Wide-field kinematics from the Mitchell Spectrograph and long-slit instruments constrain the relative contributions of stars and dark matter and increase the accuracy and precision of $\ml$ and $\mbh$.

In principle, the orbit modeling code can accommodate any dark matter density profile.  We have restricted our experiments to two functional forms: the cored LOG profile
\citep[described in, e.g.,][]{mcconnell11a},
or the NFW profile \citep{NFW}.  The free parameters in the LOG profile are the asymptotic circular speed $v_c$ and the core radius $r_c$.  The free parameters in the NFW profile are the concentration parameter $c$ and the scale radius $r_s$.  \citet{Thom07} used long-slit kinematics to determine the best-fitting LOG and NFW profiles for NGC 4889.  Our models of NGC 4889 approximate their best-fitting LOG profile.

The literature does not provide a robust estimate of the dark matter halo profile for NGC 3842, and our kinematics from the Mitchell Spectrograph extend to only $0.9\reff$.  We have tested LOG profiles with a few different values of $v_c$, as the total halo mass is proportional to $v_c^2$.  After determining the best value of $v_c$ ($350 \kms$, with $r_c = 8.0$ kpc), we tested a comparable NFW halo, scaled to enclose the same mass within the outer radius of our kinematic data.
We required $c$ and $r_s$ to follow the relation 
\begin{equation}
r_s^3 = \left( \frac{3 \times 10^{13} \; \msun}{200 \frac{4\pi}{3} \rho_{\rm crit} \, c^3} \right) 10^{\frac{1}{0.15} \left( 1.05 - \rm log_{10} \it c \right)}
\end{equation}
\citep{Rix97,mcconnell11a}, where $\rho_{\rm crit}$ is the present-day critical density.  
The resulting NFW halo has $c = 13.5$ and $r_s = 31.2$ kpc, and fits our data to the same confidence level as the LOG profile ($\Delta\chi^2_{\rm min} = 0.4$).  
Even though the density of the NFW profile increases toward $r = 0$, the total mass near the center is dominated by stars and the supermassive black hole. 
We further describe our results modeling different dark matter halos in Section~\ref{sec:BH3842}.

Our Mitchell Spectrograph kinematics extend to $1.2\reff$ for NGC 7768 and $1.5\reff$ for NGC 2832, but in both cases fail to tightly constrain the dark matter profile.  
In Section~\ref{sec:BH7768} we describe our treatment of dark matter in NGC 7768 and note that our measurement of $\mbh$ is insensitive to the dark matter profile.  in Section~\ref{sec:BH2832} we describe our dark matter models for NGC 2832 and how they affect our measurement of $\mbh$.

\subsection{Black Hole Masses and Mass-to-Light Ratios}
\label{sec:BHres}

As in \cite{mcconnell11a}, we have determined confidence intervals in $\mbh$ and $\mlr$ by running numerous models and integrating the relative likelihood function, $P \propto
e^{-\frac{1}{2}(\chi^2-\chi^2_{\rm min})}$.  Results for individual galaxies are discussed below.

\subsubsection{NGC 4889}
\label{sec:BH4889}

\begin{figure*}[!t]
 \centering
  \epsfig{figure=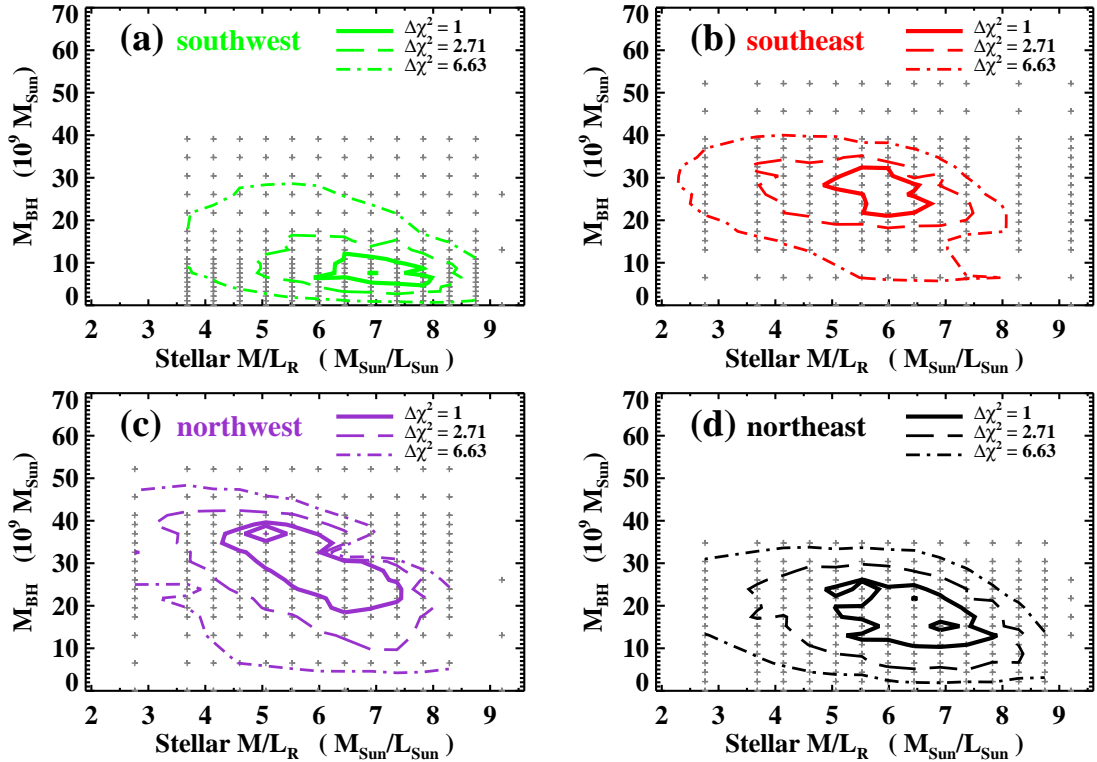,width=6.0in}
 \caption{Contours of $\chi^2$ versus $\mlr$ and $\mbh$, for models of NGC 4889, fitting LOSVDs from the GMOS IFS and from \citet{Loubser}.  Data from each quadrant of the galaxy were fit independently, and the respective results are shown in panels (a) through (d).  Contours of $\Delta\chi^2 = 1$, 2.71, and 6.63 represent confidence levels of $68\%$, $90\%$, and $99\%$ for one free parameter.  Small crosses denote individual models.  
 }
\label{fig:chi2_4889}
\vspace{0.2in}
\end{figure*}

In order to address the dramatic asymmetries in the stellar kinematics near the center of NGC 4889, we have independently run models to fit each observed quadrant of the galaxy, assuming the same central LOSVD ($r < 0.25''$) for all quadrants.  For each quadrant, we 
fit IFS kinematics from GMOS and long-slit (WHT) kinematics from \citet{Loubser}.  The resulting 
contours 
in $\chi^2 (\mbh, \mlr)$ are illustrated in Figure~\ref{fig:chi2_4889}.  There is a large degree of overlap between the 2-dimensional $68\%$ confidence regions for the northeast, southeast, and northwest quadrants, each of which contain a local maximum in $\sigma$.  After marginalizing over $\mlr$, these three quadrants yield mutually consistent black hole mass estimates.  In the southwest quadrant, $\sigma$ decreases nearly monotonically with increasing $r$, and the marginalized $68\%$ confidence interval $\mbh = 5.5 - 17.0 \times 10^9 \msun$ only agrees with the northeast quadrant.  
We wish to represent the constraints that all four quadrants impose upon the central black hole mass, and so we adopt the most extreme range of confidence limits, $\mbh = 5.5 - 37 \times 10^9 \msun$.  In other words, we exclude only those models whose orbit solution is an outlier in all four quadrants.  We define the best-fit black hole mass as the midpoint of the confidence interval above, such that $\mbh \approx 21 \pm 15 \times 10^9 \msun$.  The stellar mass-to-light ratios from all four quadrants agree within $68\%$ confidence limits.  The extreme $68\%$ confidence interval is $\mlr = 4.2 - 7.6 \mlrsun$.

For a torus of stars in NGC 4889, 
we can estimate the total enclosed mass as $M_{\rm enc} = r v^2 / G$.  
We consider a characteristic velocity, $v$, of $450 \kms$, based on the double-peaked residual features from LOSVDs near the major axis, and an
average radius, $r$, of $1.4''$ (700 pc).  These estimates yield $M_{\rm enc} = 3.3
\times 10^{10} \msun$.  Our stellar orbit models of NGC 4889 find
$\mlr$ in the range of $4.2 - 7.6 \mlrsun$, corresponding to
stellar masses of $0.7 - 1.2 \times 10^{10} \msun$ within $1.4''$.
The remaining mass of $2.1 - 2.6 \times 10^{10} \msun$ lies
near the middle of our adopted confidence interval for $\mbh$.

Although NGC 4889 does not show strong evidence for an off-center black hole, we wish to estimate the range of black hole masses that could reside at the location of the global maximum in $\sigma$.  We have fit models to the LOSVDs from the east side of NGC 4889, spatially offset such that the center of the axisymmetric gravitational potential is aligned with the LOSVD at $r = 1.4''$ along the major axis.  These models produce a $68\%$ confidence interval $\mbh = 8.7 - 24 \times 10^9 \msun$, which falls entirely within the range spanned by the models of individual quadrants.  Although these ``recentered'' models do not accurately reflect the spatial relationship between the stellar mass distribution and the observed kinematics, they support our overall conclusion that a black hole of at least $5.5 \times 10^9 \msun$ is required to produce the observed line-of-sight velocity dispersions near $400 \kms$.  Although the observed line-of-sight velocity dispersion decreases toward the center of NGC 4889, the best-fitting stellar orbit model for each quadrant predicts
that the 3-dimensional velocity dispersion $\sigma_{\rm 3D}$ rises toward the black hole.  
A non-isotropic orbital distribution, biased toward tangential orbits, causes the projected velocity dispersion to decrease even as $\sigma_{\rm 3D}$ increases.  This is further discussed in Section~\ref{sec:orbits} and illustrated in
Figure~\ref{fig:sig3d}.

\subsubsection{NGC 3842}
\label{sec:BH3842}

\begin{figure*}[htbp]
 \centering
  \epsfig{figure=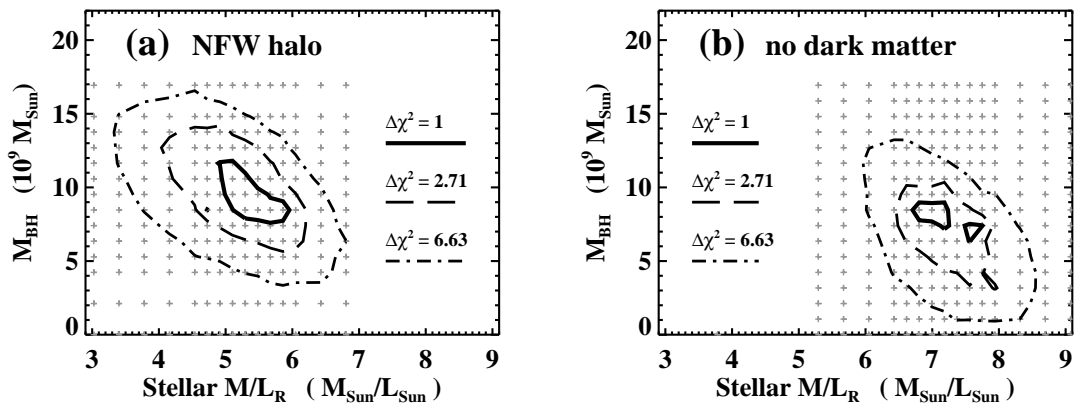,width=6.0in}
 \caption{Contours of $\chi^2$ versus $\mlr$ and $\mbh$, for models of NGC 3842, fitting LOSVDs from GMOS and the Mitchell Spectrograph.  (a) Results for our best-fitting NFW dark matter profile ($c = 13.5, r_s = 31.2$ kpc).  (b) Results for models without dark matter.  Contours of $\Delta\chi^2 = 1$, 2.71, and 6.63 represent confidence levels of $68\%$, $90\%$, and $99\%$ for one free parameter.  Small crosses denote individual models.  
}
\label{fig:chi2_3842}
\vspace{0.2in}
\end{figure*}

Models using our best estimate of the dark matter halo in NGC 3842 yield $68\%$ confidence intervals of $\mbh =  7.2 - 12.7 \times 10^9 \msun$ and $\mlr = 4.4 - 5.8 \mlrsun$.  The confidence interval for $\mbh$ overlaps with the confidence intervals from all other trials, even models with no dark matter.  The best-fit value of $\mbh$ decreases by $26\%$ when dark matter is omitted and increases by $46\%$ for our most massive halo.  
We observe a similar level of sensitivity in $\mlr$, where the most extreme trials yield best-fit values of 3.7 and $7.2 \mlrsun$.  Figure~\ref{fig:chi2_3842} illustrates the $\chi^2$ contours for models with our best-fitting NFW profile and with no dark matter.

Before acquiring Mitchell Spectrograph data, we modeled NGC 3842 with GMOS IFS data plus major-axis kinematics from \citet{Loubser}, measured using ISIS at WHT.  The more recent Mitchell Spectrograph

%
\begin{table*}[!t]
\begin{center}
\caption{Axisymmetric models of NGC 4889}
\label{tab:res4889}
\begin{tabular}[b]{ccccclccc}  
\hline
Data & $v_c$ & $r_c$ & $M_{\rm h}$(20 kpc) & $f_{\rm DM}$($r_{\rm eff}$) & $\mbh$  & $\mlr$ & $\chi^2_{\rm min}$ & $N_{\rm dof}$\\
& (km s$^{-1}$) & (kpc) & ($10^{11} \; \msun$) & & ($10^9 \; \msun$) & ($\mlrsun$) &  & \\
\\
(1) & (2) & (3) & (4) & (5) & (6) & (7) & (8) & (9)\\
\hline 
\\
GMOS southwest + WHT & 425 & 8.0 & 7.21 & 0.54 & $9.8^{+7.2}_{-4.3}$ & $6.6^{+1.0}_{-1.3}$ & 176.6 & 345 \\ 
GMOS southeast + WHT & 425 & 8.0 & 7.21 & 0.57 & $26^{+6}_{-6}$ & $5.6^{+1.1}_{-1.4}$ & 137.0 & 420 \\ 
GMOS northwest + WHT & 425 & 8.0 & 7.21 & 0.56 & $27^{+10}_{-9}$ & $5.8^{+1.2}_{-1.4}$ & 170.7 & 345 \\ 
GMOS northeast + WHT & 425 & 8.0 & 7.21 & 0.55 & $17^{+8}_{-7}$ & $6.1^{+1.2}_{-1.5}$ & 119.4 & 420 \\ 
\\
GMOS west + WHT & 425 & 8.0 & 7.21 & 0.54 & $12^{+8}_{-5.5}$ & $6.4^{+1.0}_{-1.2}$ & 180.4 & 330 \\ 
GMOS east + WHT & 425 & 8.0 & 7.21 & 0.58 & $29^{+5}_{-8}$ & $5.4^{+1.0}_{-0.9}$ & 217.4 & 442 \\ 
\\
GMOS recentered + WHT & 425 & 8.0 & 7.21 & 0.54 & $15^{+9}_{-6.3}$ & $6.5^{+0.3}_{-1.1}$ & 239.5 & 442 \\ 
\\
\hline
\end{tabular}
\end{center}
\begin{small}
\textbf{Notes:}  Column 1: Data sets included in trial.   Column 2: circular velocity of LOG dark matter halo.  Column 3: core radius of LOG dark matter halo.  For NGC 4889, we have approximated the best-fitting halo from \citet{Thom07}.  Column 4: enclosed halo mass within 20 kpc.  The ``no dark matter'' case has $v_c = 0.01$ km s$^{-1}$, $r_c = 20.0$ kpc, and $M_{\rm h} \sim 200 \msun$.  Column 5: dark matter fraction within one effective radius, assuming the best-fit values of $\mbh$ and $\mlr$.  Column 6: best-fit black hole mass.  Errors correspond to 68\% confidence intervals.  Column 7: best-fit $R$-band stellar mass-to-light ratio.  Quoted errors correspond to 68\% confidence intervals.  Column 8: minimum $\chi^2$ value for all models.  Column 9: degrees of freedom in model fits to LOSVDs.  Computed values include a smoothing factor of 1 degree of freedom per 2 velocity bins for non-parametric LOSVDs from GMOS.
\end{small}
\end{table*}
%

%

\noindent data yield slightly larger values of $\mbh$, but are consistent within $68\%$ confidence limits.  We prefer using the Mitchell Spectrograph data because they extend to larger radii ($35.3''$ versus $20.8''$) and provide full two-dimensional spatial sampling.

%
\begin{table*}[!t]
\begin{center}
\caption{Axisymmetric models of NGC 3842}
\label{tab:res3842}
\begin{tabular}[b]{ccccccclccc}  
\hline
Data & $v_c$ & $r_c$ & $c$ & $r_s$ & $M_{\rm h}$(20 kpc) & $f_{\rm DM}$($r_{\rm eff}$) & $\mbh$  & $\mlr$ & $\chi^2_{\rm min}$ & $N_{\rm dof}$\\
& (km s$^{-1}$) & (kpc) & & (kpc) & ($10^{11} \; \msun$) & & ($10^9 \; \msun$) & ($\mlrsun$) &  & \\
\\
(1) & (2) & (3) & (4) & (5) & (6) & (7) & (8) & (9) & (10) & (11)\\
\hline 
\\
GMOS + Mitchell & 0.01 & 20.0 & & & $2 \times 10^{-9}$ & 0 & $7.2^{+2.1}_{-2.9}$ & $7.2^{+0.6}_{-0.5}$ & 92.6 & 345 \\ 
GMOS + Mitchell & 350 & 8.0 & & & 4.89 & 0.51 & $10.2^{+2.9}_{-2.7}$ & $5.4^{+0.6}_{-0.7}$ & 89.3 & 345 \\ 
GMOS + Mitchell & & & 13.5 & 31.2 & 5.09 & 0.53 & $9.7^{+3.0}_{-2.5}$ & $5.1^{+0.7}_{-0.7}$ & 88.9 & 345 \\ 
GMOS + Mitchell & 500 & 8.0 & & & 9.98 & 0.75 & $13.2^{+2.2}_{-2.2}$ & $3.7^{+0.4}_{-0.6}$ & 96.2 & 345 \\ 
\\
GMOS + OSIRIS + Mitchell  & 0.01 & 20.0 & & & $2 \times 10^{-9}$ & 0 & $6.3^{+1.8}_{-1.9}$ & $7.1^{+0.6}_{-0.5}$ & 129.6 & 368 \\ 
GMOS + OSIRIS + Mitchell  & 500 & 8.0 & & & 9.98 & 0.74 & $10.2^{+1.5}_{-1.5}$ & $4.0^{+0.4}_{-0.5}$ & 138.4 & 368 \\ 
\\
GMOS + WHT & 0.01 & 20.0 & & & $2 \times 10^{-9}$ & 0 & $6.8^{+1.9}_{-2.6}$ & $7.3^{+0.5}_{-0.6}$ & 355.6 & 525 \\ 
GMOS + WHT & 500 & 8.0 & & & 9.98 & 0.71 & $9.9^{+2.1}_{-2.2}$ & $4.8^{+0.5}_{-0.8}$ & 349.3 & 525 \\ 
\\
\hline
\end{tabular}
\end{center}
\begin{small}
  \textbf{Notes:} 
Column 1: Data sets included in trial.   Column 2: circular velocity of LOG dark matter halo.  Column 3: core radius of LOG dark matter halo.  Column 4: concentration parameter for NFW dark matter halo.  Column 5: scale radius for NFW dark matter halo.  Column 6: enclosed halo mass within 20 kpc.  
Our best-fitting halo is an NFW profile with $c = 13.5$ and $r_s = 31.2$ kpc.  The equivalent LOG halo ($v_c = 350 \kms$; $r_c = 8.0$ kpc) fits our data to the same level of confidence ($\Delta\chi^2_{\rm min} = 0.4$).  
The ``no dark matter'' case has $v_c = 0.01$ km s$^{-1}$, $r_c = 20.0$ kpc, and $M_{\rm h} \sim 200 \msun$.  
Omitting dark matter from the models causes $\mbh$ to decrease by $26\%$, consistent within errors.
Column 7: dark matter fraction within one effective radius, assuming the best-fit values of $\mbh$ and $\mlr$.  Column 8: best-fit black hole mass.  Errors correspond to 68\% confidence intervals.  Column 9: best-fit $R$-band stellar mass-to-light ratio.  Quoted errors correspond to 68\% confidence intervals.  Column 10: minimum $\chi^2$ value for all models.  Column 11: degrees of freedom in model fits to LOSVDs.  Computed values include a smoothing factor of 1 degree of freedom per 2 velocity bins for non-parametric LOSVDs from GMOS and OSIRIS.
\end{small}
\end{table*}
%

%

OSIRIS and GMOS provide independent measurements of stellar kinematics in NGC 3842, for $r \leq 0.7''$.  In spite of the potential for higher spatial resolution, systematic contaminants in our OSIRIS spectra force us to increase $S/N$ by binning the data to identical spatial scales as for GMOS.  We have run orbit models fitting LOSVDs from OSIRIS
and GMOS simultaneously (as well as Mitchell Spectrograph data at large
radii). Including the OSIRIS data causes the best-fit value of $\mbh$ to
decrease by up to 23\%, likely because OSIRIS data show a less drastic
increase in $\sigma$ than data from GMOS.  
For the best-fit models, including OSIRIS data produces higher average $\chi^2$ values
per LOSVD, even after ignoring the central regions where LOSVDs
from OSIRIS and GMOS are not fully consistent. 
The kinematic fits to OSIRIS data may have
systematic errors, resulting from imperfect sky subtraction and difficulty defining the $H$-band continuum.  Consequently, we judge the models of NGC 3842 with only GMOS and Mitchell Spectrograph data to be the most reliable.

%
\begin{table*}[!t]
\begin{center}
\caption{Axisymmetric models of NGC 7768}
\label{tab:res7768}
\begin{tabular}[b]{ccccclccc}  
\hline
Data & $v_c$ & $r_c$ & $M_{\rm h}$(20 kpc) & $f_{\rm DM}$($r_{\rm eff}$) & $\mbh$  & $\mlr$ & $\chi^2_{\rm min}$ & $N_{\rm dof}$\\
& (km s$^{-1}$) & (kpc) & ($10^{11} \; \msun$) & & ($10^9 \; \msun$) & ($\mlrsun$) &  & \\
\\
(1) & (2) & (3) & (4) & (5) & (6) & (7) & (8) & (9)\\
\hline  
\\
OSIRIS + Mitchell  + WHT & 0.01 & 20.0 & $2 \times 10^{-9}$ & 0 & $1.3^{+0.5}_{-0.4}$ & $6.1^{+0.7}_{-0.8}$ & 77.4 & 285 \\ 
OSIRIS + Mitchell  + WHT & 350 & 8.0 & 4.89 & 0.32 & $1.2^{+0.5}_{-0.4}$ & $5.2^{+0.5}_{-0.6}$ & 78.8 & 285 \\ 
\\
\hline
\end{tabular}
\end{center}
\begin{small}
\textbf{Notes:}  See Table~\ref{tab:res4889} for notes and definitions.  For WHT measurements, we only include five LOSVDs, covering radii between $1''$ and $3''$.
The LOSVDs extracted from the Mitchell Spectrograph cover five radial bins between $ 3.0''$ and $28.5''$ (0.1 - 1.2 $\times \reff$).   Several models with different dark matter profiles, including no dark matter and the dark matter model listed here, all have $\Delta\chi^2 < 1$.  
Our best-fit black hole mass for NGC 7768 is insensitive to the assumed dark matter profile, a testament to the strength of our OSIRIS data.
\end{small}
\end{table*}
%

%
\begin{figure*}[!t]
 \centering
  \epsfig{figure=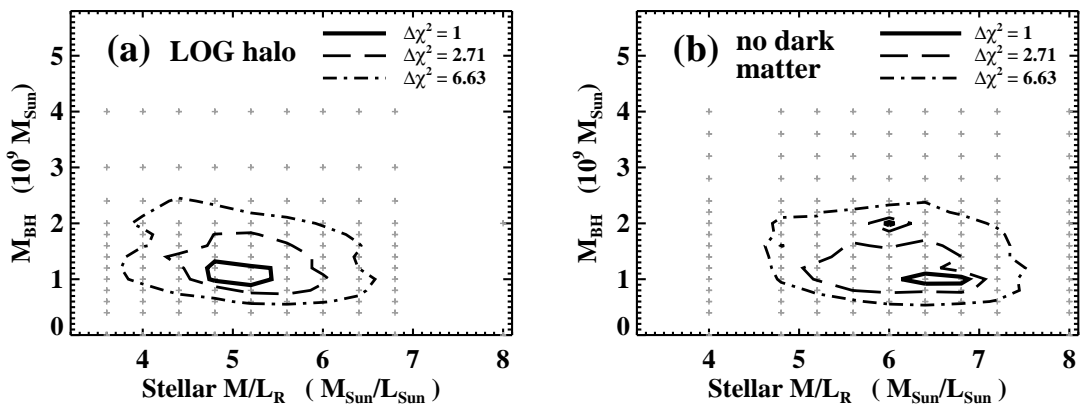,width=6.0in}
 \caption{Contours of $\chi^2$ versus $\mlr$ and $\mbh$, for models of NGC 7768, fitting LOSVDs from OSIRIS and the Mitchell Spectrograph, plus five long-slit data points from \citet{Loubser}.  (a) Results for a LOG dark matter profile ($v_c = 350 \kms, r_c = 8.0$ kpc).  (b) Results for models without dark matter.  Contours of $\Delta\chi^2 = 1$, 2.71, and 6.63 represent confidence levels of $68\%$, $90\%$, and $99\%$ for one free parameter.  Small crosses denote individual models.  
 }
\label{fig:chi2_7768}
\vspace{0.2in}
\end{figure*}

\subsubsection{NGC 7768}
\label{sec:BH7768}

For NGC 7768, we have combined IFU data from OSIRIS and the Mitchell Spectrograph, plus five long-slit measurements at radii of 1-3$''$.  We have run two trials with full sampling of $\mbh$ and $\ml$.  One trial includes a LOG halo profile, matching our best-fitting approximation for NGC 3842 ($v_c = 350 \kms; r_c = 8.0$ kpc).  The other trial does not include dark matter.  The black hole masses from the two trials are in excellent agreement, and the best-fit values of $\mlr$ agree within $68\%$ confidence limits.   The trial with no dark matter actually yields a lower value of $\chi^2_{\rm min}$, albeit with marginal significance ($\Delta\chi^2_{\rm min} = 1.4$).  Considering the lower value of $\chi^2_{\min}$ from the trial without dark matter, we formally adopt a black hole mass of $1.3^{+0.5}_{-0.4} \times 10^9 \msun$.  The range of $\mlr$ from both trials is 4.6-6.8 $\mlrsun$ ($68\%$ confidence).  The respective $\chi^2$ contours are shown in Figure~\ref{fig:chi2_7768}.

Given our initial trials' consistent measurements of $\mbh$ and $\ml$, we have not run finely sampled trials for additional dark matter profiles.  In order to check whether any dark matter profile for NGC 7768 can fit our kinematics as well as the trial without dark matter, we have run a grid of models 
coarsely sampling 
$\ml$, $v_c$, and $r_c$.  This trial yields several models whose $\chi^2$ values are consistent with $\chi^2_{\rm min}$ for the trial without dark matter.

\subsubsection{NGC 2832}
\label{sec:BH2832}

%
\begin{table*}[!t]
\begin{center}
\caption{Axisymmetric models of NGC 2832}
\label{tab:res2832}
\begin{tabular}[b]{ccccclccc}  
\hline
Data & $v_c$ & $r_c$ & $M_{\rm h}$(20 kpc) & $f_{\rm DM}$($r_{\rm eff}$) & $\mbh$  & $\mlr$ & $\chi^2_{\rm min}$ & $N_{\rm dof}$\\
& (km s$^{-1}$) & (kpc) & ($10^{11} \; \msun$) & & ($10^9 \; \msun$) & ($\mlrsun$) &  & \\
\\
(1) & (2) & (3) & (4) & (5) & (6) & (7) & (8) & (9)\\
\hline  
\\
GMOS IFU + Mitchell + GMOS slit & 350 & 32.0 & 1.59 & 0.28 & $ < 5.2$ & $8.1^{+0.3}_{-0.5}$ & 129.0 & 495 \\ 
GMOS IFU + Mitchell + GMOS slit & 560 & 40.0 & 2.90 & 0.47 & $5.9^{+3.1}_{-3.2}$ & $7.3^{+0.4}_{-0.4}$ & 125.1 & 495 \\ 
\\
GMOS IFU + Mitchell & 560 & 40.0 & 2.90 & 0.47 & $7.2^{+3.0}_{-3.1}$ & $7.2^{+0.3}_{-0.5}$ & 115.2 & 450 \\  
\\
\hline
\end{tabular}
\end{center}
\begin{small}
\textbf{Notes:}  See Table~\ref{tab:res4889} for notes and definitions.  For GMOS slit measurements, we only include three LOSVDs, covering radii between $4''$ and $7''$.
\end{small}
\end{table*}
%

%
\begin{figure*}[!t]
 \centering
  \epsfig{figure=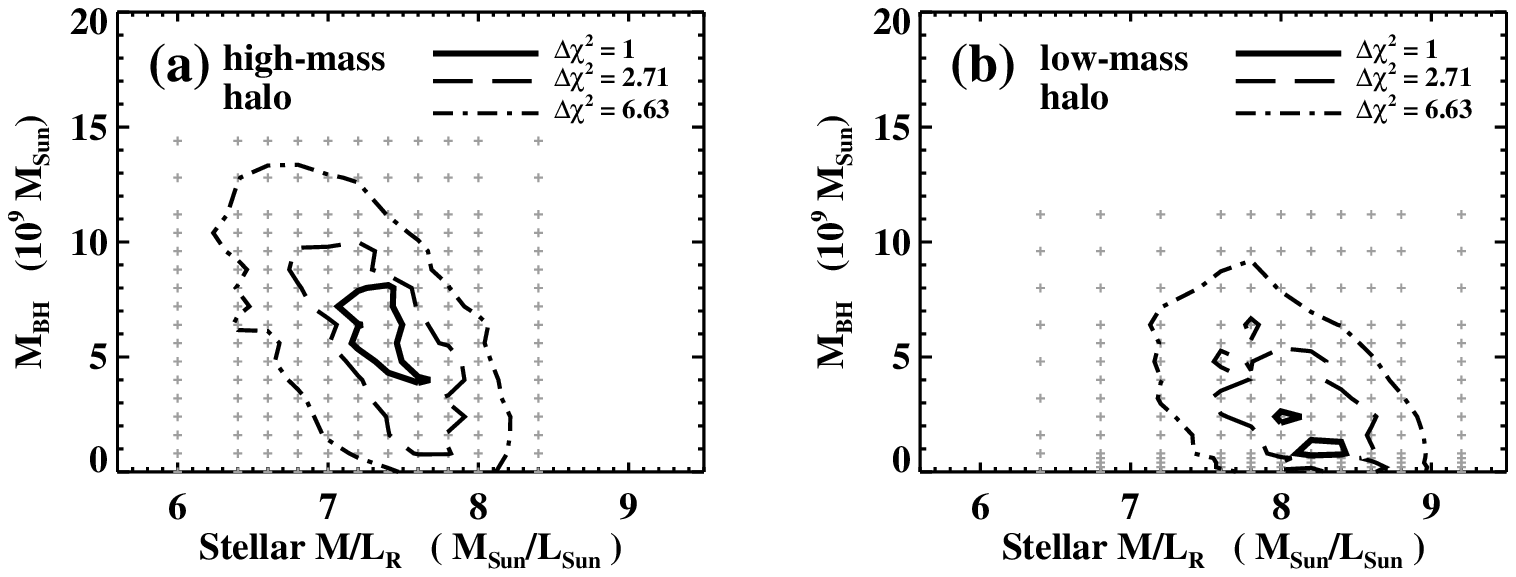,width=6.0in}
 \caption{Contours of $\chi^2$ versus $\mlr$ and $\mbh$, for models of NGC 2832, fitting LOSVDs from GMOS and Mitchell Spectrograph IFS data, plus three long-slit data points from \citet{Loubser}.  (a) Results for a high-mass LOG dark matter profile ($v_c = 560 \kms, r_c = 40.0$ kpc).  (b) Results for a lower-mass LOG profile ($v_c = 350 \kms, r_c = 32.0$ kpc).  Contours of $\Delta\chi^2 = 1$, 2.71, and 6.63 represent confidence levels of $68\%$, $90\%$, and $99\%$ for one free parameter.  Small crosses denote individual models.  
 }
\label{fig:chi2_2832}
\vspace{0.2in}
\end{figure*}
%

%
\begin{figure}[!b]
\vspace{-0.1in}
 \centering
  \epsfig{figure=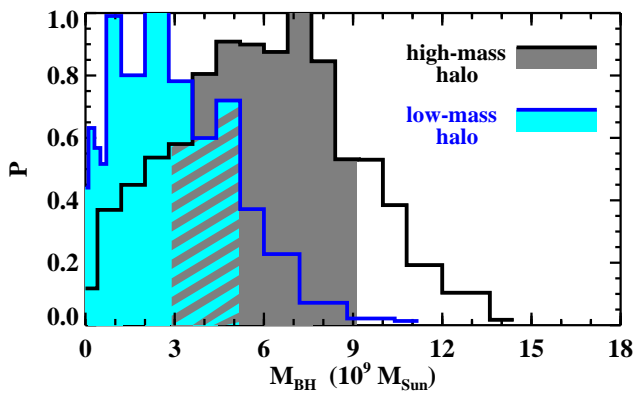,width=3.5in}
 \caption{Likelihood, $P$, versus $\mbh$ for models of NGC 2832, after marginalizing with respect to $\mlr$.   The black line represents the trial with a high-mass LOG dark matter profile ($v_c = 560 \kms, r_c = 40.0$ kpc), and the grey shaded region represents the corresponding 68\% confidence interval.  The blue line and cyan shaded region represent the trial with a lower-mass LOG profile ($v_c = 350 \kms, r_c = 32.0$ kpc).  The diagonally hashed region is where the two confidence intervals overlap.  Both likelihood distributions have been normalized to a maximum value of 1.  Considering both trials, and discrepancies between different instruments' kinematic measurements for NGC 2832, we adopt an upper limit of $\mbh = 9 \times 10^9 \msun$.}
\label{fig:lik2832}
\vspace{0.1in}
\end{figure}

For NGC 2832, we have combined IFU data from GMOS and the Mitchell Spectrograph, along with three long-slit measurements at radii of 4-$7''$, and have run full trials varying $\mbh$ and $\ml$ for two dark matter halos.  Figure~\ref{fig:chi2_2832} illustrates the $\chi^2$ contours, and Figure~\ref{fig:lik2832} shows the likelihood distribution of $\mbh$ from each trial, after marginalizing with respect to $\ml$.  Models for the less massive halo ($v_c = 350$ km s$^{-1}$; $r_c = 32.0$ kpc) yield a best-fit black hole mass of $2.9 \times 10^9 \msun$ and a $68\%$ upper confidence limit of $5.2 \times 10^9\msun$.  However, the likelihood distribution falls slowly as $\mbh$ approaches zero, and has sufficient noise between $\mbh = 0$ and $\mbh = 6.0 \times 10^9 \msun$ to cast doubt upon any lower limit in $\mbh$.  In contrast, the more massive halo ($v_c = 560$ km s$^{-1}$; $r_c = 40.0$ kpc) yields a best-fit black hole mass of $5.9 \times 10^9 \msun$, and the corresponding likelihood distribution declines cleanly to both sides, producing a $68\%$ confidence interval of $\mbh = 2.7 - 9.0 \times 10^9 \msun$.  We obtain consistent results for this halo when the three long-slit data points at intermediate radii are excluded.

When data from GMOS and the Mitchell Spectrograph are modeled simultaneously, the minimum $\chi^2$ values for all input $\mbh$ and $\ml$ significantly favor the more massive halo ($\Delta\chi^2_{\rm min} = 3.9$).  However, there are multiple reasons to be skeptical of the corresponding confidence limits for $\mbh$.  First, the sharp jump between the velocity dispersions measured with GMOS and those measured with the Mitchell Spectrograph suggests that systematic errors in the kinematic fitting could be responsible for the apparent signature of a massive black hole.  Second, the best-fit stellar mass-to-light ratio of $7.3 \mlrsun$ exceeds the upper limits for $\mlr$ in NGC 4889 and NGC 3842.
Third, the best-fit value of $\mbh$ in NGC 2832 is sensitive to the dark matter halo profile, and one trial with a plausible halo fails to produce a significant black hole detection.  
Given the possibility of systematic errors in the kinematics, we do not find convincing evidence for a specific dark matter profile in NGC 2832.  
We conservatively interpret our models of NGC 2832 to provide an upper limit of $\mbh = 9.0 \times 10^9 \msun$, to $68\%$ confidence.

\subsection{Tangentially Biased Orbits}
\label{sec:orbits}

\begin{figure*}[!t]
\centering
  \epsfig{figure=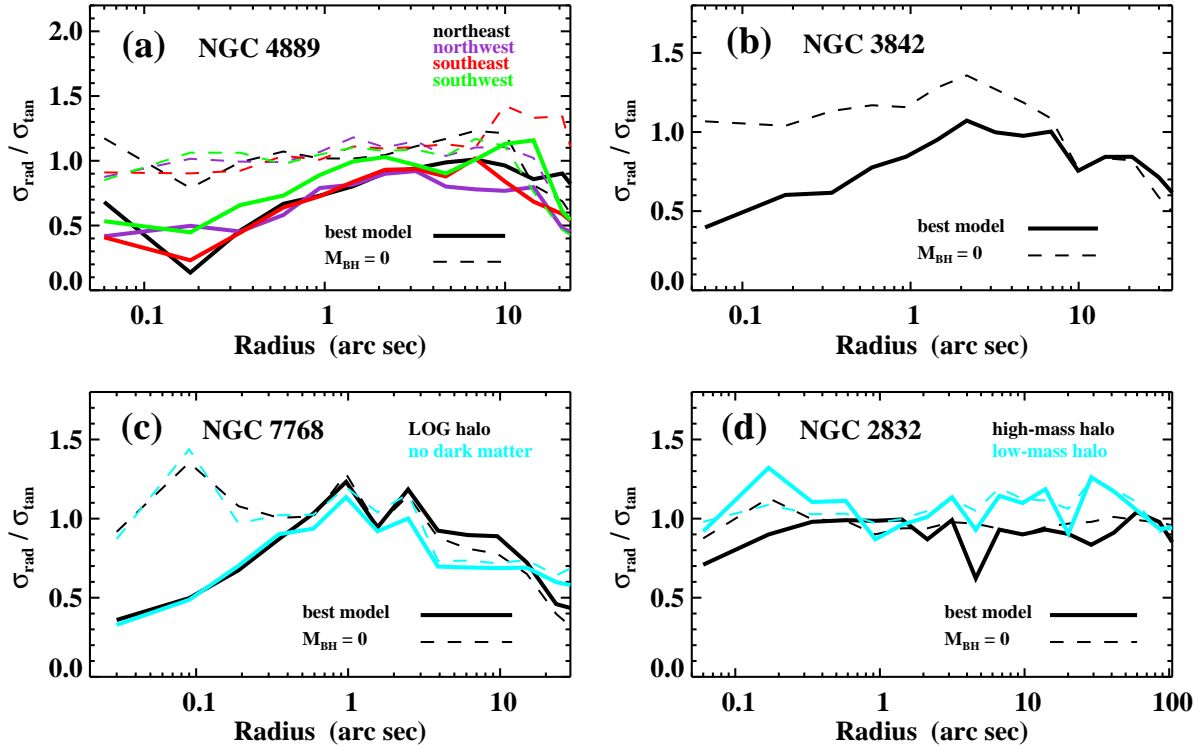,width=6.3in}
 \caption{Anisotropy of model stellar orbits in each galaxy, as a function of radius.  Solid lines represent the best-fitting models for each set of LOSVDs, and dashed lines represent the best-fitting models without a black hole.  We define $\sigma_{\rm rad}$ and $\sigma_{\rm tan}$ such that their ratio is unity for an isotropic orbit distribution.  (a) NGC 4889.  Different colors represent the northeast (black; $\mbh = 1.2 \times 10^{10} \msun$), northwest (purple; $\mbh = 2.6 \times 10^{10} \msun$), southeast (red; $\mbh = 2.2 \times 10^{10} \msun$), and southwest (green; $\mbh = 6.0 \times 10^9 \msun$) quadrants.   (b) NGC 3842.  The best fitting model has $c = 13.5$, $r_s = 31.2$ kpc, $\mbh = 8.0 \times 10^9 \msun$.  (c) NGC 7768.  The best-fitting model with dark matter (black lines) has $v_c = 350 \kms$, $r_c = 8.0$ kpc, and $\mbh = 1.0 \times 10^9 \msun$.  The best-fitting model without dark matter (cyan lines) has $\mbh = 1.0 \times 10^9 \msun$.  (d) NGC 2832.  The best-fitting model for the high-mass dark matter halo (black lines) has $v_c =  560 \kms$, $r_c = 40.0$ kpc, and $\mbh = 7.2 \times 10^9 \msun$.  The best-fitting model for the low-mass halo (cyan lines) has $v_c =  350 \kms$, $r_c = 32.0$ kpc, and $\mbh = 8.0 \times 10^8 \msun$.}
\vspace{0.1in}
\label{fig:bias}
\end{figure*}

In NGC 4889, NGC 3842, and NGC 7768, the supermassive black hole
coincides with a deficiency of radial orbits near the galaxy
center.  In Figure~\ref{fig:bias}, we plot orbital
anisotropy versus radius for each BCG and compare predictions from our orbital models
  with the best-fit black hole mass and without a black hole.
Specifically, we examine the ratio of the radial and tangential velocity
dispersions in the 3-dimensional models.  The tangential dispersion term
includes azimuthal rotation and is defined as $\sigma^2_{\rm tan} \equiv
\frac{1}{2}(\sigma_\theta^2 + \sigma_\phi^2 + v_\phi^2)$,
such that $\sigma_{\rm rad} / \sigma_{\rm tan} = 1$ for an isotropic orbit distribution.  

Figure~\ref{fig:bias} illustrates that for NGC 4889, NGC 3842, and NGC
  7768, the models with black holes predict strong tangential velocity
  biases toward the galactic centers (solid curves), while the models
  without black holes predict fairly isotropic velocity distributions with
  little radial dependence (dashed curves).  After averaging over the four
quadrants using the best-fit black hole masses, the minimum value of
$\sigma_{\rm rad}/\sigma_{\rm tan}$ in NGC 4889 is approximately 0.3.  The best models for NGC 3842 and NGC 7768 have central values of $\sigma_{\rm rad}/\sigma_{\rm tan} \approx 0.3 - 0.4$. This
degree of tangential bias is comparable to the central values in a handful
of early-type galaxies \citep[e.g.,][]{Geb03,Geb07,SG10}.  NGC 4889 and NGC
3842, however, are exceptional in terms of the size of the central regions
dominated by tangential orbits.  In both galaxies, this region extends to
approximately $2''$, or 1 kpc.  By comparison, tangential bias extends to
only 50-200 pc throughout a broad sample of early-type galaxies with black
hole mass measurements (Gebhardt et al. 2003; 2007; Nowak et al. 2008; Shen
\& Gebhardt 2010; Gebhardt et al. 2011; c.f., G\"{u}ltekin et al. 2009b).
In NGC 7768, we measure a tangential bias for $r < 0.5''$ (270 pc) for
  a black hole mass of $1\times 10^9 M_\odot$.

The scarcity of radial orbits at the centers of massive galaxies may be a consequence of past core-scouring by one or more pairs of in-spiraling black holes.  $N$-body simulations of merging galaxies and black holes indicate that stars on radial orbits are more likely to interact with a binary black hole, resulting in their ejection from the galaxy center \citep[e.g.,][]{QH97,Milos01}.  
Supporting this picture, \citet{Geb03} ran stellar orbit models of twelve galaxies with black holes and noted that the four core-profile galaxies had a higher degree of tangential bias than the eight power law galaxies.
Photometric data of NGC 4889 show a stellar core with a radius of 750 pc \citep{Lauer07}, similar to the the scale on which the tangential bias appears.  
In the sample of 118 core-profile galaxies from \citet{Lauer07}, only eight exhibit core radii greater of 750 pc or greater; all eight are BCGs or cD galaxies of comparable luminosity.
In NGC 3842, the core radius is 310 pc, roughly one third the size of the tangentially biased region.
The surface brightness profile for NGC 7768 turns over at approximately 220 pc, similar to the size of the tangentially biased region.  

\begin{figure}[!h]
\centering
\vspace{-0.2in}
  \epsfig{figure=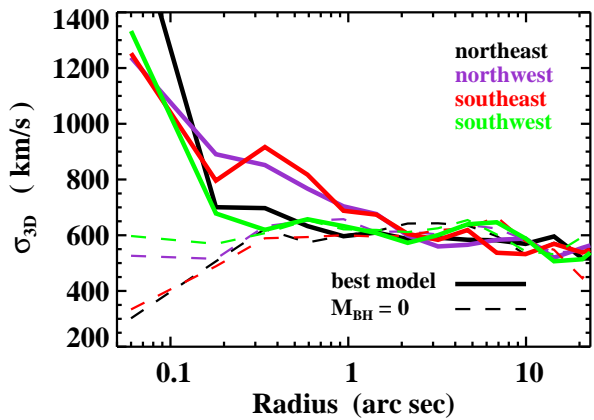,width=3.5in}
 \caption{Three-dimensional velocity dispersion, $\sigma_{\rm 3D}$, 
as a function of radius predicted by our orbital models for NGC 4889.  Solid lines represent the best-fitting model for each quadrant, and dashed lines represent the best-fitting models without a black hole.  Different colors represent the northeast (black; $\mbh = 1.2 \times 10^{10} \msun$), northwest (purple; $\mbh = 2.6 \times 10^{10} \msun$), southeast (red; $\mbh = 2.2 \times 10^{10} \msun$), and southwest (green; $\mbh = 6.0 \times 10^9 \msun$) quadrants.}
\label{fig:sig3d}
\vspace{0.1in}
\end{figure}

It is important to note that the decrease in velocity dispersions
  towards the black hole, as illustrated in Figure~\ref{fig:kin4889} for NGC 4889, occurs for the
  observed \textit{line-of-sight} velocity dispersions.  The 3-dimensional
  velocity dispersion $\sigma_{\rm 3D}$ is not directly measureable but is
  predicted by our orbital models.  Figure~\ref{fig:sig3d} illustrates the
  radial trends in $\sigma_{\rm 3D}$ for models of each quadrant of NGC
  4889.  As expectd, $\sigma_{\rm 3D}$ rises towards the center of the
  galaxy when a black hole is included in the model.
  Figures~\ref{fig:bias} and \ref{fig:sig3d} together show that a
  non-isotropic orbital distribution, biased toward tangential orbits, can
  cause the light-of-sight velocity dispersion to decrease even as
  $\sigma_{\rm 3D}$ increases.

Our models of NGC 4889 indicate that orbital anisotropy is responsible for the local minimum in
$\sigma$ at the galaxy center.  In NGC 3842, the increase in $\sigma$ is
less pronounced than would be observed for isotropic orbits about a
$10^{10}$-$M_\odot$ black hole.  For both galaxies, models with $\mbh = 0$
have enough freedom to imitate the observed kinematics, but require a
nearly isotropic orbital distribution to do so
(Figure~\ref{fig:bias}a,~\ref{fig:bias}b).  In NGC 7768, a radial bias is
required to reproduce the sharp increase in $\sigma$
(Figure~\ref{fig:bias}c).  Although the LOSVDs generated by models with and
without black holes are qualitatively similar, the models with $\mbh = 0$ produce higher values of $\chi^2$ in every trial.
 
 NGC 2832 has nearly isotropic orbits at all radii, and the orbital discrepancy between models with and without a black hole is weaker than in NGC 4889 and NGC 3842.  This is particularly true for models with the less massive dark matter halo, where we lack a confident black hole detection.  The best-fitting model with the more massive halo has $\mbh = 7.2 \times 10^9 \msun$, and $\sigma_{\rm rad}/\sigma_{\rm tan}$ declines to 0.7 over the the inner $0.3''$ (150 pc).

%

%

\subsection{Possible Systematic Errors}
\label{sec:moreerror}

The common practice for reporting errors in stellar dynamical measurements of $\mbh$ and $\ml$ is to report the range of input $\mbh$ and $\ml$ values that fall within a specific confidence interval for a given set of models.  Yet modeling a galaxy requires explicit assumptions about its stellar content and the structure of its gravitational potential.  Systematic errors from these assumptions will likely increase the overall measurement errors for $\mbh$ and $\ml$.  

For each BCG, our models assume an edge-on inclination.  This is indirectly
supported by the observed axis ratios of 0.68 to 0.86, which imply
relatively eccentric intrinsic shapes even for an edge-on system.  Models
with more face-on inclinations could yield higher black hole masses.  We
also model each galaxy as an oblate axisymmetric ellipsoid, in contrast to
the presence of isophotal twists in NGC 3842 and NGC 2832 (Table~\ref{tab:sample}) and
circumstantial evidence that some BCGs are prolate or triaxial
\citep{PSH91,RLP93,BKMaQ06}.  In an early experiment with triaxial models,
the best-fit black hole mass for NGC 3379 increased by a factor of two,
  whereas the black hole mass for M32 was unchanged from axisymmetric models \citep{vdB10}.  The increase for NGC 3379 may have arisen solely from a
change in the best-fit inclination (R.C.E. van den Bosch, 2011, private
communication).

Massive elliptical galaxies are known to exhibit spatial gradients in metallicity \citep[e.g.,][]{FFI95, KA99, Mehlert03, Brough07,Greene12}, and gradients in stellar age and $\alpha$-element enhancement have been reported in some cases \citep{FFI95, CGA10}. 

Massive ellipticals could exhibit corresponding gradients in $\ml$, yet our orbit models assume that $\ml$ is constant.  
This assumption could bias our estimates of the enclosed stellar mass at large or small radii, depending upon the steepness of the gradient and the radial sampling of our kinematic data.  
Preliminary single-burst stellar population modeling of M87 indicates that $\ml$ decreases 
by $\sim 50\%$ from $r = 0$ to $r \sim 2\reff$
(Graves \& Murphy, 2012, in prep).  
Most of our kinematics for the BCGs herein correspond to $r < \reff$, and we estimate that a gradient in $\ml$ would bias our measurement of the central $\ml$ by $< 10\%$.  
Kinematic and photometric data probe the galaxy's total mass-to-light ratio, and models with dark matter have more flexibility to reproduce each galaxy's total mass profile.  Trade-offs between dark matter and gradients in $\ml$ could mitigate potential biases in our measurements of $\mbh$.  
Still, our knowledge of the individual mass components in BCGs would be improved by using stellar line indices and population models to measure gradients in $\ml$ independently. 

A fundamental assumption of all orbit superposition models is that the stellar motions reflect a steady-state gravitational potential.  The strongly asymmetric kinematics in NGC 4889 present the possibility of an unrelaxed stellar system, although we do not witness corresponding photometric irregularities.  Even with our conservative treatment of the asymmetric data, our models would be unable to assess the central black hole mass if the kinematics of NGC 4889 reflected a temporary condition.

%
\section{Conclusion and Discussion}
\label{sec:disc}

%
\begin{figure*}[!t]
\centering
  \epsfig{figure=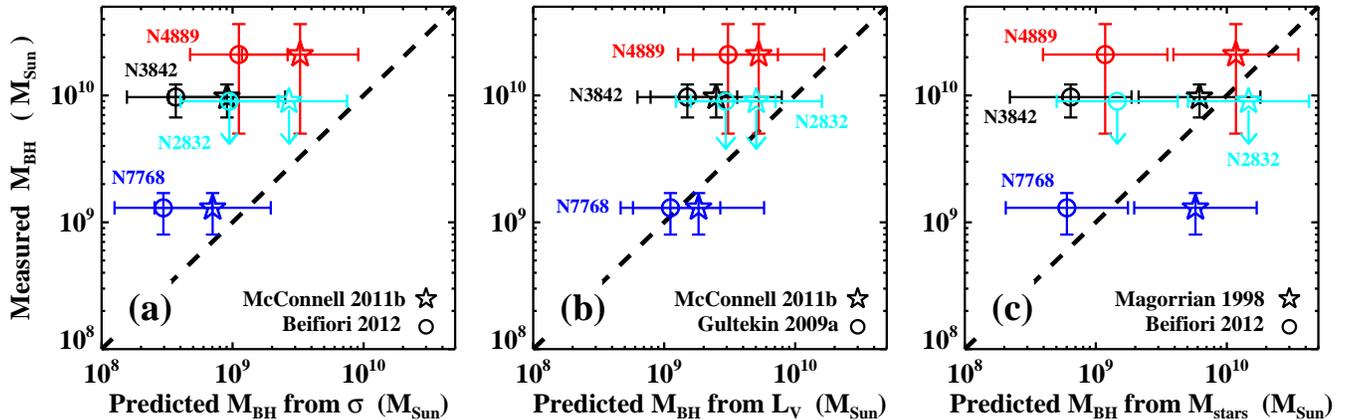,width=7.0in}
 \caption{Measured black hole masses from this work versus predicted black hole masses from various scaling relations.  (a) Predicted masses from the $\mbh-\sigma$ relation, log($\mbh$) = $\alpha + \beta$log($\sigma/ 200 \kms$).  Stars represent the relation of McConnell et al. (2011b; $\alpha = 8.29$, $\beta = 5.12$, $\epsilon_0 = 0.43$), and circles represent Beifiori et al. (2012; $\alpha = 7.99$, $\beta = 4.42$, $\epsilon_0 = 0.36$).  (b) Predicted masses from the $\mbh-L_V$ relation, log($\mbh$) = $\alpha + \beta$log($L_V/ 10^{11} L_{\odot,V}$).  Stars represent the relation of McConnell et al. (2011b; $\alpha = 9.16$, $\beta = 1.16$, $\epsilon_0 = 0.50$), and circles represent G\"{u}ltekin et al. (2009; $\alpha = 8.95$, $\beta = 1.11$, $\epsilon_0 = 0.38$).  (c) Predicted masses from the $\mbh-M_\star$ relation, log($\mbh$) = $\alpha + \beta$log($M_\star/ 10^{11} \msun$).  Stars represent the relation of Magorrian et al. (1998; $\alpha = 8.77$, $\beta = 0.96$), and circles represent Beifiori et al. (2012; $\alpha = 7.84$, $\beta = 0.91$, $\epsilon_0 = 0.46$).  The scaling relations for each panel were selected to span the widest range of predicted black hole masses.  The horizontal error bars in each panel are dominated by the intrinsic scatter, $\epsilon_0$, in log($\mbh$), as estimated for each relation.  \citet{Magorrian} do not provide an estimate of $\epsilon_0$ with respect to $M_\star$, so we have adopted $\epsilon_0 = 0.46$ from \citet{Beifiori12} in panel (c).  Horizontal error bars also contain a small contribution from measurement errors in $\sigma$, $L_V$, or $M_\star$.  Vertical error bars represent measurement errors in $\mbh$, as determined by our stellar orbit models.}
\vspace{0.2in}
\label{fig:Mcompare}
\end{figure*}

We have reported three black hole masses (NGC 4889, NGC 3842, NGC 7768) and
one upper limit on $\mbh$ (NGC 2832) in four of the nearby Universe's most
massive galaxies at distances of $\sim 100$ Mpc.  As the basis for our
analysis, we have presented high-resolution, 2-dimensional measurements of
stellar kinematics using GMOS at Gemini Observatory North and OSIRIS at
Keck Observatory.  For three of our galaxies, we have used wide-field
stellar kinematics from the Mitchell Spectrograph at McDonald Observatory.  

We have derived black hole masses from stellar kinematics using orbit
superposition models.  We find $\mbh = 2.1 \pm 1.6 \times 10^{10} \msun$ for NGC 4889, $\mbh = 9.7^{+3.0}_{-2.5} \times 10^9 \msun$ for NGC 3842, $\mbh = 1.3^{+0.5}_{-0.4} \times 10^9 \msun$ for NGC 7768, and $\mbh < 9.0 \times 10^9 \msun$ for NGC 2832.  The orbit models also determine the stellar mass-to-light ratio $\mlr$, which we convert to $\mlv$ as described in Section~\ref{sec:phot}.
We find $\mlr = 5.9 \pm 1.7 \mlrsun$ ($\mlv = 7.4 \pm 2.1 \mlvsun$) for NGC 4889, $\mlr = 5.2 \pm 0.8 \mlrsun$ ($\mlv = 7.1 \pm 1.1 \mlvsun$) for NGC 3842, $\mlr = 5.7 \pm 1.1 \mlrsun$ ($\mlv = 8.8 \pm 1.7 \mlvsun$) for NGC 7768, and $\mlr = 7.6^{+0.8}_{-0.7} \mlrsun$ ($\mlv = 9.7 \pm 1.0 \mlvsun$) for NGC 2832.
  
The orbital distributions from the best-fitting models indicate that NGC
4889 and NGC 3842 are depleted of radial orbits, out to radii of
approximately 1 kpc. These extended tangential bias regions have similar
sizes to the galaxies' photometric cores, and may be linked to the extremely
massive black hole ($\mbh \sim 10^{10} \msun$) in each galaxy.  NGC 7768
has a black hole mass of only $\sim 10^9 \msun$ and a correspondingly
modest tangential bias region.

Several correlations between black hole mass and scalar galaxy properties have been explored by numerous authors.  Below we focus on the $\mbh - \sigma$ relation, the $V$-band $\mbh - L$ relation, and the relation between $\mbh$ and bulge stellar mass, $M_\star$.   
We estimate $M_\star$ for each BCG by multiplying $L_V$ by $\mlv$.  
 
 In Figure~\ref{fig:Mcompare}, we compare our measured values of $\mbh$ in NGC 4889, NGC 3842, and NGC 7768, and our upper limit for NGC 2832, to predicted black hole masses from the $\mbh-\sigma$, $\mbh-L_V$, and $\mbh-M_\star$ relations.  
In particular, we display the $\sigma$- and $L$-based predictions from \citet{mcconnell11b}, who use the most up-to-date sample of directly measured black hole masses.  
The scaling relations derived from this sample are steepened by the large black hole masses in NGC 4889 and NGC 3842, as well as upward revisions of $\mbh$ in M87 and M60 \citep{GT09,SG10,Geb11}.  The $\mbh-\sigma$ relation is additionally steepened by the inclusion of eight late-type galaxies with maser-based measurements of $\mbh$ \citep{Kondratko,Greene10,Kuo11}. 
As a result, the \citet{mcconnell11b} relations predict the largest values of $\mbh$ in BCGs.
For comparison, we have considered a large number of reported power-law fits to earlier galaxy samples and have 
determined which fits yield
the lowest predicted values of $\mbh$ for our BCGs.  These minimum predictions are also displayed in Figure~\ref{fig:Mcompare}, in order to illustrate the full range of investigations to date.
For the $\mbh-M_\star$ relation, we display the predicted black hole masses from \citet{Magorrian} and \citet{Beifiori12}, which span the full range of $M_\star$-based predictions.
Along with the $\mbh$ values predicted from the mean relations, we consider the intrinsic scatter, $\epsilon_0$, in log($\mbh$) at fixed $\sigma$, $L_V$, or $M_\star$.  The horizontal error bars in Figure~\ref{fig:Mcompare} essentially illustrate $\epsilon_0$; measurement errors in $\sigma$, $L_V$, and $M_\star$ have much smaller effects on the predicted black hole masses.  
 
The left panel of Figure~\ref{fig:Mcompare} indicates that our
measurements of $\mbh$ in NGC 4889, NGC 3842, and NGC 7768 are all greater
than the predicted values from various fits to the $\mbh-\sigma$ relation.
For NGC 4889 and NGC 3842, the discrepancy exceeds the intrinsic scatter in
$\mbh$ regardless of which fit to $\mbh-\sigma$ we select.  For NGC 7768,
$\mbh$ exceeds the mean $\mbh-\sigma$ relations from \citet{Gultekin}, \citet{Graham11} and
\citet{Beifiori12} by more than the corresponding estimates of
$\epsilon_0$, but is consistent with the power-law fit and $\epsilon_0$
estimate from \citet{mcconnell11b}.
 
The predicted
black hole masses from $\mbh-L_V$, indicated in the middle panel of Figure~\ref{fig:Mcompare}, are uniformly more massive than the
$\sigma$-based predictions.
Our large measurement errors for $\mbh$ in NGC 4889, combined with
estimates of $\epsilon_0$, yield agreement between the measured black hole
mass and the $\mbh-L_V$ relation of \citet{mcconnell11b}.  However, the
black hole in NGC 3842 is significantly more massive than predicted by any
version of the $\mbh-L_V$ relation.  For NGC 7768, our measurement of
$\mbh$ agrees with all versions of the $\mbh-L_V$ relation.

The right panel of Figure~\ref{fig:Mcompare} shows that the $\mbh-M_\star$
relation offers the widest range of predicted black hole masses from the
literature.  Only the historical relation from \citet{Magorrian} is
consistent with the measured black hole masses of both NGC 4889 and NGC
3842.  More recent estimates of $\mbh-M_\star$ by \citet{Sani11} and
\citet{Beifiori12}, however, predict black hole masses an order of
magnitude lower and are consistent with our measurement in NGC 7768.

Our upper limit for $\mbh$ in NGC 2832 is consistent with the full range of
predicted black hole masses from $\sigma$, $L_V$, and $M_\star$.

There has been recent dispute over the existence of a fundamental correlation between $\mbh$ and the Virial mass or circular velocity of the host galaxy's dark matter halo 
(e.g., Ferrarese 2002; Kormendy \& Bender 2011; Volonteri et al. 2011; also Beifiori et al. 2012 and references therein).
We can compare NGC 4889 and NGC 3842 to this correlation by using $v_c$ from their best-fitting LOG dark matter halos, or by computing the Virial mass of the best-fitting NFW halo for NGC 3842 ($M_{200} = 8.5 \times 10^{12} \msun$).  Using either metric, we find that the black holes in NGC 4889 and NGC 3842 are several times more massive than those in galaxies with comparable dark matter halos.

The present-day black hole mass function is a testable prediction of galaxy evolution and feedback models.  In particular, 
different models make divergent predictions for the number density of black holes with $\mbh \sim 10^9 - 10^{10} \msun$ 
\citep[see, e.g.,][and references therein]{KM12}.
Our measurements of $\mbh \sim 10^{10} \msun$ in NGC 4889 and NGC 3842, plus the $6.3 \times 10^9 \msun$ black hole in M87 \citep{Geb11} and the $4.7 \times 10^9 \msun$ black hole in M60 \citep{SG10}, place a lower limit of four galaxies with 
$\mbh > 10^{9.5} \msun$
within a local volume $\sim 10^6$ Mpc$^3$.  
This is slightly higher than the space density predicted from models of proportional black hole and spheroid growth in major mergers \citep[e.g.,][]{Hopkins08}. 
Several other BCGs lie within 100 Mpc, and the discovery of more black holes with 
$\mbh > 10^{9.5} \msun$
could approach the predicted space density from models by \citet{Shen09}, in which all quasars follow a universal light curve and the most massive black holes accrete gas until redshifts $\sim 1$.
\citet{NT09} have predicted the black hole mass function from the X-ray luminosity function of active galactic nuclei (AGN).  Our lower limit is consistent with their prediction using the original luminosity function from \citet{Ueda03} but does not support their hypothesis of a truncated luminosity function for AGN with $\mbh > 10^9$.
Models by \citet{Yoo07} find that black hole-black hole mergers can increase $\mbh$ by factors $\sim 2$ in galaxy clusters, with the largest black holes reaching $1-1.5 \times 10^{10} \msun$.  Black hole mergers also dominate the growth of the largest black holes in recent models by \citet{Fanidakis11}, but their prescriptions for quasar-mode (near-Eddington) and radio-mode (low-Eddington) accretion predict relatively low space densities for black holes with $\mbh > 10^9 \msun$.

The steep $L-\sigma$ relation for BCGs may arise from extreme size evolution, either via low-angular momentum mergers \citep[e.g.,][]{BKMaQ05,BKMaQ06} or via cannibalism of numerous low-density systems \citep[e.g.,][]{OT75,OH77,Hopkins10,Oser10}.  In both scenarios, size evolution is driven by gas-poor or ``dry'' processes.  Dry major mergers should preserve any initial relation between $\mbh$ and stellar mass.  For black hole growth to mirror stellar mass growth in the cannibalism scenario, the devoured satellites must indeed contain central black holes, and dynamical friction must deliver them efficiently to the center of the BCG.  
The forms of the $\mbh-\sigma$, $\mbh-L$, and $\mbh-M_\star$ relations for BCGs could provide additional insight to their progenitors and growth mechanisms. 

Our BCGs are not the only systems that have been observed to deviate from the black hole scaling relations.
The $\mbh-\sigma$, $\mbh-L$, and $\mbh-M_\star$ relations each have outliers that span a variety of galaxy masses and environments \citep[e.g.,][]{Nowak08,Greene10,Nowak10,Reines11,Bogdan12}.  
In fact, some merging models predict that the most massive galaxies will exhibit smaller deviations from the mean black hole scaling relations, as a consequence of experiencing greater numbers of hierarchical mergers \citep{Peng07,VN09,JM11}.  In this scenario, black hole growth and galaxy growth need not be coupled through direct feedback processes.  
On the other hand, systematic offsets for $\mbh$ in BCGs could point to unique evolutionary processes near cluster centers.  A larger sample of BCGs with measured $\mbh$ is necessary to distinguish between systematic trends and random scatter.

We aim to use high-resolution and wide-field kinematic data of several more galaxies to explore the high-mass slopes of the empirical correlations between galaxies and supermassive black holes, and to quantify the total and intrinsic scatter in $\mbh$ for massive galaxies in different environments.  This investigation and future studies are important tests for the hypothesis of universal scaling relations between black holes and their host galaxies, and for galaxy evolution models using different modes and timescales for black hole accretion.

\medskip
We thank Jenny Greene, Remco van den Bosch, and Jonelle Walsh for helpful discussions during this investigation, and our referee for highlighting a number of complementary investigations.  NJM thanks Anne-Marie Weijmans for instrument training at McDonald Observatory.
This work was supported by NSF AST-1009663.  Support for C.-P.M. is provided in part by the Miller Institute for Basic Research in Science, University of California, Berkeley.  KG acknowledges support from NSF-0908639.  The W. M. Keck Observatory is operated as a scientific partnership among the California Institute of Technology, the University of California, and the National Aeronautics and Space Administration. The Observatory was made possible by the generous financial support of the W. M. Keck Foundation.  The Gemini Observatory is operated by the 
Association of Universities for Research in Astronomy, Inc., under a cooperative agreement 
with the NSF on behalf of the Gemini partnership: the National Science Foundation (United 
States), the Science and Technology Facilities Council (United Kingdom), the 
National Research Council (Canada), CONICYT (Chile), the Australian Research Council (Australia), 
Minist\'{e}rio da Ci\^{e}ncia e Tecnologia (Brazil) 
and Ministerio de Ciencia, Tecnolog\'{i}a e Innovaci\'{o}n Productiva (Argentina).  GMOS data were obtained under Program GN-2003A-Q-11. 
We thank the Cynthia and George Mitchell Foundation for funding the Mitchell Spectrograph, formerly known as VIRUS-P.
Stellar orbit models were run on supercomputers at the Texas Advanced Computing Center (TACC).  
In addition, we wish to acknowledge the significant cultural role and reverence that the summit of Mauna Kea has always had within the indigenous Hawaiian community.  We are most fortunate to have the opportunity to conduct observations from this mountain.

\end{document}